\documentclass{article}
\usepackage[T1]{fontenc}
\usepackage[utf8]{inputenc}
\usepackage{arxiv}
\usepackage{amssymb}
\usepackage{amsmath}
\usepackage{txfonts}
\usepackage{graphicx}
\usepackage[numbers]{natbib}
\usepackage{threeparttable}
\usepackage{booktabs}
\usepackage{array}
\usepackage{tabularx}
\usepackage{placeins}
\usepackage{stfloats}
\usepackage{microtype}
\graphicspath{{figs/}}
\usepackage[colorlinks,allcolors=blue,bookmarksnumbered]{hyperref}
\usepackage{orcidlink}
\hypersetup{
  pdfauthor={Zhaoyu Chen, Zili Chen, Jingwen Xu, Yonghua Ding, Wei Jiang, Donghui Xia, Ya Zhang},
  pdftitle={Phasor-particle-in-cell algorithm for bidirectional external-circuit coupling of inductively coupled plasma}}
\renewcommand{\shorttitle}{Bidirectional phasor--PIC external-circuit coupling}
\date{}
\makeatletter
\newcommand{\figw}[1]{\if@twocolumn\columnwidth\else#1\columnwidth\fi}
\makeatother

\let\preprintappendix\appendix
\renewcommand{\appendix}{%
  \preprintappendix
  \setcounter{equation}{0}%
  \renewcommand{\theequation}{\Alph{section}.\arabic{equation}}%
  \renewcommand{\theHequation}{appendix.\Alph{section}.\arabic{equation}}%
  \renewcommand{\thefigure}{\Alph{section}.\arabic{figure}}%
  \renewcommand{\thetable}{\Alph{section}.\arabic{table}}%
}

\title{Phasor-particle-in-cell algorithm for bidirectional external-circuit coupling of inductively coupled plasma}

\author{%
\begin{minipage}{0.96\textwidth}
\centering\normalfont
Zhaoyu Chen\textsuperscript{a}\,\orcidlink{0000-0003-3388-4212},
Zili Chen\textsuperscript{a}\,\orcidlink{0000-0002-2104-5369},
Jingwen Xu\textsuperscript{b}\,\orcidlink{0000-0003-1385-3320},
Yonghua Ding\textsuperscript{a}\,\orcidlink{0000-0001-7475-6802},\\
Wei Jiang\textsuperscript{c}\,\orcidlink{0000-0002-9394-585X},
Donghui Xia\textsuperscript{a,*}\,\orcidlink{0000-0002-5325-7622},
Ya Zhang\textsuperscript{b,*}\,\orcidlink{0000-0003-0473-467X}\\[6pt]
{\small
\textsuperscript{a}State Key Laboratory of Advanced Electromagnetic Technology,
International Joint Research Laboratory of Magnetic Confinement Fusion and Plasma Physics,
School of Electrical and Electronic Engineering, Huazhong University of Science and Technology,
Wuhan 430074, China\\[3pt]
\textsuperscript{b}School of Physics and Mechanics, Wuhan University of Technology,
Wuhan 430070, China\\[3pt]
\textsuperscript{c}School of Physics, Huazhong University of Science and Technology,
Wuhan 430074, China\\[4pt]
\textsuperscript{*}Corresponding authors:
\href{mailto:xiadh@hust.edu.cn}{xiadh@hust.edu.cn} (Donghui Xia);
\href{mailto:yazhang@whut.edu.cn}{yazhang@whut.edu.cn} (Ya Zhang).
}
\end{minipage}%
}

\begin{document}
\raggedbottom
\maketitle

\begin{abstract}
We develop a self-consistent phasor--particle-in-cell algorithm to capture
bidirectional inductive and capacitive feedback between an inductively
coupled plasma and a distributed radio-frequency coil circuit. The method couples a
two-dimensional axisymmetric particle-in-cell model with Monte Carlo
collisions to a network that resolves each turn's current and potential.
Precomputed unit-current Helmholtz solutions provide the vacuum impedance
matrix, and projection of the plasma-only field supplies the series back
electromotive force. An established charge-based electrostatic coupling is
extended to individual turns: surface charges obtained from a discrete Gauss law
consistent with the Poisson solver yield shunt displacement currents through
their fundamental harmonics. Both responses are accumulated over a complete
radio-frequency period to update the relaxed circuit state once per period.
Vacuum tests gave differences of $-1.58$\% in single-solenoid inductance
and $-4.38$\% in mutual inductance relative to analytical references. A
particle-free five-node manufactured benchmark with both channels active
gave a scaled complex-state error of $1.90\times10^{-5}$ against the
continuous reference and approximately second-order spatial convergence.
Application to an unshielded argon reference cell reached statistically
stationary states at coil-port powers of 60 and 100~W. Net capacitive heating
accounted for 29.35\% and 20.07\%, respectively, of the combined inductive
and capacitive plasma power.
Independently accumulated field- and circuit-side powers, including
conductor loss, agreed within about 0.5\%.
These results support turn-resolved bidirectional field--circuit coupling
for axisymmetric plasma simulations with a fundamental-frequency circuit
representation.
\end{abstract}


\section{Introduction}
\label{sec:intro}

An inductively coupled plasma (ICP) and its radio-frequency (RF) supply form
an electrically coupled field--plasma--circuit system. The coil currents
determine the induced azimuthal electric field, whereas the turn potentials
determine the electrostatic field and capacitive coupling through the
surrounding space. Capacitive transfer from an ICP coil through a dielectric
window measurably changes the RF plasma potential~\cite{watanabe1999radio}.
Measurements also show that antenna voltage and current, conductor loss, and
plasma-absorbed power depend on the loaded circuit state
\cite{singh2008electrical,godyak2017power}. Circuit-coupled ICP models
accordingly predict that the matching network can alter mode transitions and
the time-dependent partition of delivered power
\cite{xu2015equivalent,qu2020power}.

Kinetic simulation adds a second layer of coupling because the electrical
load must be obtained from noisy, time-dependent particle quantities.
Implicit particle formulations were developed to relax the fast-scale
restrictions of explicit PIC integration
\cite{brackbill1982implicit,langdon1983direct}. Recent PIC developments
include energy- and charge-conserving implicit schemes for axisymmetric
collisional plasmas~\cite{angus2024axisymmetric}, unstructured
body-of-revolution electromagnetic solvers~\cite{na2024borpic}, and
domain-decomposition Poisson solvers~\cite{mao2025poisson}. Fully implicit
electromagnetic, alternating-direction implicit, and Darwin formulations have
since enabled multidimensional kinetic simulations of ICPs
\cite{mattei2017fully,fu2024kinetic,sydorenko2025darwin}; the low-frequency
Darwin approximation itself removes propagating light-wave modes while
retaining electrostatic and inductive fields~\cite{hewett1994low}. In these
kinetic ICP implementations, however, the RF coil is driven by a specified
current rather than by a distributed external-circuit solution. Earlier
axisymmetric PIC/MCC studies from our group likewise used prescribed or
power-controlled coil
excitation~\cite{chen2025eh,chen2026rf}; in
Ref.~\cite{chen2026rf}, the coil-current amplitude was adjusted to meet a
target inductive power deposition without solving the external coil
equations.

Self-consistent circuit coupling has been developed most directly for
electrostatic discharges. An early method solved the potential of a bounded
one-dimensional PIC plasma and a series RLC circuit simultaneously, using
electrode charge conservation~\cite{verboncoeur1993simultaneous}.
An energy- and charge-conserving implicit formulation subsequently
incorporated an external RLC network into the field iterations for
collisional bounded plasmas~\cite{eremin2022bounded}.
Matching networks have also been coupled to reduced plasma circuit
models~\cite{schmidt2018consistent}. These formulations
establish the importance of nonlinear electrode--circuit feedback, but their
driven plasma boundary is electrostatic and does not provide the self- and
mutual-inductive ports of a multi-turn ICP coil.

For ICPs, transformer-equivalent models provide a useful global description
of the reflected plasma resistance, inductance change, and power transfer
\cite{elfayoumi1998electromagnetic,godyak2017power}. Distributed conductor
models retain more of the antenna geometry. The partial-element equivalent
circuit (PEEC) formulation maps multiconductor fields to partial inductive and
capacitive elements~\cite{ruehli1974equivalent}, while a multiconductor
transmission-line model has combined mutual-inductance and capacitance
matrices to reproduce the measured input impedance of a large-area ICP
antenna~\cite{guittienne2017electromagnetic}. The latter uses measured or
parameterized plasma properties in the antenna model rather than a kinetic
plasma response computed from the same spatial solution.

The computational problem addressed here is therefore to feed both spatial
plasma responses back to a distributed RF coil circuit without duplicating
the vacuum inductance or capacitance contained in the simulation domain. We
formulate a cycle-coupled phasor--PIC algorithm for this purpose. Unit-current
field solutions provide the per-turn vacuum
impedance matrix, while the plasma-only azimuthal field is projected as a
series back electromotive force. Independently, the charge-based
electrostatic coupling used in the underlying two-dimensional PIC/MCC model
\cite{chen2024gec} is extended to the individual coil turns: a discrete Gauss
law using the same face coefficients as the Poisson solver provides each
turn's free charge, and its fundamental harmonic gives the corresponding
shunt displacement current. These two feedback quantities close a
$(2N+1)$-unknown distributed network once per RF period.

The present calculations use two approximations appropriate to the intended
coil and operating regime. The conductor resistance is estimated from the
frequency-dependent skin depth and the length of each turn using the local
round-wire model in Eq.~\eqref{eq:skin}. This estimate does not resolve
winding-dependent current redistribution within the conductors, including
proximity effects from neighboring turns~\cite{chen2022winding}. Such
corrections may be needed for larger planar coils; they are neglected for the
compact coil considered here. The inductive plasma response is retained only
at the fundamental frequency because our earlier ICP study found higher
harmonics of the plasma current and induced field to be negligible under the
conditions studied~\cite{chen2025eh}. This motivates the present truncation,
while the frequency-domain formulation permits extension to additional
harmonics. Our recent current waveform tailoring (CWT) study illustrates
their use in controlling electron kinetics through nonsinusoidal coil
excitation~\cite{chen2026cwt}.

The main contributions of this work are: (i) a common source--projection
construction for the vacuum self- and mutual-inductive impedances and the
plasma reaction voltages; (ii) extension of an established charge-based
electrostatic coupling to turn-resolved conductors and integration of the
resulting displacement currents into the distributed circuit; (iii) a
partitioned RF-cycle update that combines the two feedback
channels with relaxation, power control, and independent circuit-side and
energy-balance diagnostics; and (iv) staged verification of the self-
and mutual-inductive responses followed by a five-node manufactured test
of the joint inductive--capacitive network, field extraction, and
period-delayed feedback. The joint test excludes particles and quantifies
the spatial-discretization errors of the coupled field--circuit solution.
The complete algorithm is then applied to a two-dimensional GEC reference
cell. Section~\ref{sec:methodology}
presents the formulation and coupling sequence, Sec.~\ref{sec:results}
reports the verification tests and the plasma application, and
Sec.~\ref{sec:conclusions} summarizes the main findings and limitations.

\section{Methodology}
\label{sec:methodology}


\subsection{Overview of the coupling framework}
\label{subsec:overview}

We consider an ICP sustained by an RF coil in a
two-dimensional axisymmetric $(Z,R)$ geometry, with the azimuthal direction
denoted by $\theta$. The particle advance, collision, electrostatic field,
and weighted-particle infrastructure is inherited from earlier
two-dimensional direct-implicit PIC/MCC implementations
\cite{chen2024gec,chen2025swpc,chen2025eh,chen2026rf}; the present work
focuses on the additional coil--field--circuit closure. When a PIC/MCC model
of such a discharge prescribes the
coil current as an external excitation, the
plasma load cannot modify the coil current, the potential distribution along
the coil, or the matching state. The algorithm presented here instead
determines these quantities from a distributed external-circuit solution and
closes the field--circuit loop through both coil--plasma energy-transfer
channels:

\begin{itemize}
\item \emph{Inductive channel.} The coil current produces the induced
  azimuthal field $E_\theta$ through a Helmholtz equation; the
  plasma current $\hat J_{\theta,p}$ in turn produces a reaction field that is
  projected onto each coil turn as a series back electromotive force (emf)
  $\hat V_{p,k}^{\mathrm{ind}}$.
\item \emph{Capacitive channel.} The RF potentials $\hat V_k$ of the coil
  turns are imposed as Dirichlet boundary values of the electrostatic Poisson
  equation; the resulting free charge $Q_k(t)$ on the conductor surface forms
  the displacement current $I_{\mathrm{cap},k}(t)=\mathrm{d}Q_k/\mathrm{d}t$,
  whose fundamental phasor $\hat I_{\mathrm{cap},k}=\mathrm{i}\omega\hat Q_k$
  returns to the external circuit as a per-turn shunt load.
\end{itemize}

The algorithm has a three-level time structure. At the PIC level, with time
step $\Delta t$, particles are pushed, collide, and deposit charge and
current; the Poisson equation is solved and the coil surface charge is
sampled at every step. At the RF-cycle level, every
\begin{equation}
N_{\mathrm{RF}}=\operatorname{nint}\!\left(\frac{1}{f\,\Delta t}\right)
\label{eq:nrf}
\end{equation}
PIC steps the fundamental-frequency field and circuit state are updated once,
where $f$ is the driving frequency. The input is required to satisfy the
phase-closure condition
\begin{equation}
\left|f\,\Delta t\,N_{\mathrm{RF}}-1\right|<10^{-10}
\label{eq:phaseclosure}
\end{equation}
so that the discrete Fourier projections below cover exactly one RF period.
At the macroscopic level, the plasma state and the circuit operating point
co-evolve over many RF cycles, which can be viewed as a nonlinear fixed-point
iteration with the RF period as the outer step.

The PIC state remains time resolved within each RF period, but only the
fundamental Fourier components of the azimuthal plasma current and conductor
charges enter the field and circuit updates. This is narrower than
time-domain electrostatic PIC--circuit formulations that retain nonlinear
electrical waveforms and higher harmonics
\cite{eremin2022bounded,chen2024gec}. Higher-harmonic circuit
feedback is omitted in the calculations reported here, consistent with the
scope described in Sec.~\ref{sec:intro}.

The $\exp(+\mathrm{i}\omega t)$ convention with $\omega=2\pi f$ is used
throughout, and all voltages and currents are \emph{peak} phasors. A real
quantity is written as
\begin{equation}
X(t)=X_c\cos\omega t+X_s\sin\omega t
=\operatorname{Re}\!\left(\hat X e^{\mathrm{i}\omega t}\right),
\qquad \hat X=X_c-\mathrm{i}X_s
\label{eq:phasor}
\end{equation}
The cycle-averaged complex power therefore carries the factor $1/2$,
$S=\tfrac12\hat V\hat I^{*}$. The azimuthal field and current are stored as
two real arrays per component,
\begin{equation}
E_\theta(t_n)=E_c\cos\theta_n-E_s^{(\mathrm{sto})}\sin\theta_n,
\qquad
\theta_n=\frac{2\pi\,(n\bmod N_{\mathrm{RF}})}{N_{\mathrm{RF}}}
\label{eq:et_reconstruct}
\end{equation}
where the stored sine array equals $-E_s$, so that the field phasor is
$\hat E_\theta=E_c+\mathrm{i}E_s^{(\mathrm{sto})}$. The fundamental-frequency
Fourier projection of a sampled quantity $X_n$ over one closed RF period is
\begin{equation}
X_c=\frac{2}{N_{\mathrm{RF}}}\sum_{n=1}^{N_{\mathrm{RF}}}X_n\cos\theta_n,
\qquad
X_s=\frac{2}{N_{\mathrm{RF}}}\sum_{n=1}^{N_{\mathrm{RF}}}X_n\sin\theta_n
\label{eq:fourier}
\end{equation}
This single sign convention constrains the Fourier projections, the two real
Helmholtz solves, the circuit current components, the Dirichlet node
potentials, the power definitions and the charge differentiation; any local
sign change would break the phase consistency of the whole coupling chain.

\subsection{Inductive field calculation and circuit feedback}
\label{subsec:inductive}

Phasor formulations of the azimuthal ICP field have long been used in
two-dimensional GEC models~\cite{lymberopoulos1995two}. The harmonic
Helmholtz equation and particle-current Fourier projection used here follow
our earlier axisymmetric PIC/MCC implementation
\cite{chen2025eh,chen2026rf}; the present
extension constructs turn-resolved circuit ports from that field solution.

Under axisymmetry, the azimuthal electric field satisfies
\begin{equation}
\left(
\frac{\partial^2}{\partial z^2}
+\frac{\partial^2}{\partial R^2}
+\frac{1}{R}\frac{\partial}{\partial R}
-\frac{1}{R^2}
-\mu_0\varepsilon_0\varepsilon_r\frac{\partial^2}{\partial t^2}
\right)E_\theta
=\mu_0\frac{\partial J_\theta}{\partial t}
\label{eq:helmholtz_time}
\end{equation}
Keeping only the fundamental frequency and defining
\begin{equation}
\mathcal L=
\frac{\partial^2}{\partial z^2}
+\frac{\partial^2}{\partial R^2}
+\frac{1}{R}\frac{\partial}{\partial R}
-\frac{1}{R^2}
+\mu_0\varepsilon_0\varepsilon_r\omega^2
\label{eq:helmholtz_op}
\end{equation}
the phasor form reads
\begin{equation}
\mathcal L\hat E_\theta=\mathrm{i}\omega\mu_0\hat J_\theta
\label{eq:helmholtz_phasor}
\end{equation}
Decomposing $\hat J_\theta=J_c-\mathrm{i}J_s$ and
$\hat E_\theta=E_c-\mathrm{i}E_s$ shows that the complex problem is obtained
from two real solves with the same operator,
\begin{equation}
\mathcal L E_c=\mu_0\omega J_s,
\qquad
\mathcal L(-E_s)=\mu_0\omega J_c
\label{eq:helmholtz_two_real}
\end{equation}
For the unit-current solves below, the corresponding scaled real operator is
denoted by $\mathcal K=\mathcal L/(\mu_0\omega)$. The same five-point
discretization is used for both real components. The verification tests of
Sec.~\ref{sec:results} employ the magnetoquasistatic limit of this operator
with the displacement-current term omitted, whereas the GEC application
retains the full operator (Table~\ref{tab:coverage}).
On the symmetry axis, regularity of the azimuthal field requires
\begin{equation}
E_\theta(z,0)=0
\label{eq:axis_inductive_bc}
\end{equation}
This axial regularity condition is distinct from a metallic or grounded
boundary condition. The condition $E_\theta=0$ is also imposed on the outer
truncation boundaries and on ideal metal surfaces. The coil cross-sections
enter this equation as \emph{volume current source regions}, not as
Dirichlet conductors. The discrete system is solved in parallel by a PETSc
Krylov solver.

To connect the inductive field calculation to the external circuit, let the
coil consist of $N$ turns. Turn $k$ occupies a discrete cross-section
$\Omega_k$ containing $N_k$ source nodes, with discrete area and winding sign
\begin{equation}
A_k=N_k\,\Delta z\,\Delta R,\qquad s_k\in\{-1,+1\}
\end{equation}
The unit-current shape function is
\begin{equation}
g_k(z_i,R_j)=
\begin{cases}
\dfrac{s_k}{A_k}, & (i,j)\in\Omega_k,\\[1ex]
0, & \text{otherwise},
\end{cases}
\qquad
\sum_{(i,j)\in\Omega_k}g_k(z_i,R_j)\,\Delta z\,\Delta R=s_k
\label{eq:shape}
\end{equation}
Every conductor node is required to belong to exactly one electrical turn,
and the normalization~\eqref{eq:shape} is verified at initialization. The
same $g_k$ is used both as the excitation of the field equation and as the
test function of the port-voltage projection; this
``same-source--same-test'' construction reduces non-reciprocity errors
caused by inconsistent discrete port definitions. With this normalization, a
long thin turn of length $L$ represents a cylindrical current sheet carrying
the total ampere-turn current $I_p$, i.e.\ a surface current
$K_\theta=I_p/L$.

For each turn, the unit-current response field is solved \emph{once} at
initialization (or when the derived port state is rebuilt after a restart),
\begin{equation}
\mathcal K U_k=g_k
\label{eq:unitfield}
\end{equation}
For arbitrary turn-current phasors $\hat I_k=I_{c,k}-\mathrm{i}I_{s,k}$, the
total coil field follows by linear superposition. With the storage convention
introduced above, the composed azimuthal field is
\begin{equation}
E_c=E_c^{(p)}+\sum_{k=1}^{N}I_{s,k}\,U_k,
\qquad
(-E_s)=(-E_s^{(p)})+\sum_{k=1}^{N}I_{c,k}\,U_k
\label{eq:compose}
\end{equation}
where $(E_c^{(p)},-E_s^{(p)})$ is the plasma-only field obtained from the two
solves~\eqref{eq:helmholtz_two_real} with the particle-current harmonics.
The exchange of the current components between the real and imaginary arrays
is not an empirical correction; it originates from the phase rotation of
$\partial J_\theta/\partial t$ between the cosine and sine components.

The projection of the unit field of turn $l$ onto the port of turn $k$
defines the vacuum impedance matrix
\begin{equation}
\begin{aligned}
Z_{\mathrm{vac},kl}
&=-\mathrm{i}\int_{\Omega}g_k(z,R)\,U_l(z,R)\,\mathrm{d}V,\\
&=-\mathrm{i}\sum_{i,j}g_{k,ij}\,U_{l,ij}\,2\pi R_j\,\Delta z\,\Delta R
\end{aligned}
\label{eq:zvac}
\end{equation}
with the axisymmetric volume element
$\mathrm{d}V=2\pi R\,\mathrm{d}z\,\mathrm{d}R$.
Diagonal entries contain the self inductance and off-diagonal entries the
mutual inductances, $\operatorname{Im}Z_{\mathrm{vac},kl}=\omega M_{kl}$. In a
linear, source-free vacuum the matrix should be symmetric; the relative
reciprocity error
\begin{equation}
\epsilon_{\mathrm{rec}}
=\frac{\max_{k,l}\left|Z_{\mathrm{vac},kl}-Z_{\mathrm{vac},lk}\right|}
{\max_{k,l}\left|Z_{\mathrm{vac},kl}\right|}
\label{eq:reciprocity}
\end{equation}
is reported as a built-in diagnostic. This field-derived matrix provides the
self-consistent vacuum self- and mutual-inductive contributions in the
circuit equations.

PEEC models likewise derive partial self- and mutual-inductive elements from
multiconductor fields~\cite{ruehli1974equivalent}. Equation~\eqref{eq:zvac}
is not, however, a general PEEC discretization in three dimensions. It is a
port projection of the axisymmetric Helmholtz
operator, and it supplies only the vacuum inductive matrix; capacitance inside
the resolved domain remains in the Poisson problem below.

The two Helmholtz solves with the measured plasma-current harmonics give the
plasma-only field
$\hat E_{\theta,p}=E_c^{(p)}+\mathrm{i}(-E_s^{(p)})$. The reaction emf induced
in turn $k$ and the voltage drop entering the segment KVL of the circuit are
\begin{equation}
\hat{\mathcal E}_{p\rightarrow k}
=\int_{\Omega}g_k\,\hat E_{\theta,p}\,\mathrm{d}V,
\qquad
\hat V_{p,k}^{\mathrm{ind}}=-\hat{\mathcal E}_{p\rightarrow k}
=-\int_{\Omega}g_k\,\hat E_{\theta,p}\,\mathrm{d}V
\label{eq:vpind}
\end{equation}
Only the plasma-only field enters this projection. Using the total field,
which already contains the superposed unit fields, would count the vacuum
inductance twice, once in $Z_{\mathrm{vac}}$ and once in
$\hat V_{p,k}^{\mathrm{ind}}$.

Equations~\eqref{eq:zvac} and~\eqref{eq:vpind} express the inductive port
responses as weighted spatial integrals, with the axisymmetric factor
$2\pi R$. The normalization in Eq.~\eqref{eq:shape} preserves the prescribed
total turn current when a source cross-section is represented by different
numbers of grid points. Comparisons between discretizations therefore refer
to the same physical source region and spatial fundamental-current
distribution. For fixed geometry, material coefficients and boundary
conditions, consistent representations of that distribution approximate
the same continuous field and port response, subject to field-discretization,
quadrature and solver errors.

\subsection{Electrostatic field calculation and capacitive circuit feedback}
\label{subsec:capacitive}

The electrostatic field $(E_z,E_R)$ is solved in the time domain every PIC
step using the direct implicit particle response
\cite{langdon1983direct,chen2024gec}. For the field solve at $t_n$, the
particle contribution to $\rho$ is deposited at predicted positions,
before their correction by the newly solved electrostatic field.
Deposited dielectric-surface charge is also included in the discrete source
where present. The numerical susceptibility $\chi(z,R)$ is assembled from
the same predicted particle populations and held fixed during this solve.
The total electrostatic potential $\phi$ satisfies
\begin{equation}
\nabla\cdot\left[\varepsilon_0\,\varepsilon_r(1+\chi)\,\nabla\phi\right]=-\rho
\label{eq:poisson}
\end{equation}
The role of $\chi$ follows from the linearized charge correction associated
with the implicit particle displacement~\cite{langdon1983direct}.
Writing $\mathbf E_{\mathrm{es}}^n=-\nabla\phi^n$ and the physical
displacement field as $\mathbf D_{\mathrm{es}}^n=\varepsilon_0\varepsilon_r
\mathbf E_{\mathrm{es}}^n$, the scalar closure in Eq.~\eqref{eq:poisson}
can be expressed as
\begin{equation}
\begin{aligned}
\mathbf P_{\mathrm{num}}^n
&=\varepsilon_0\varepsilon_r\chi^n\mathbf E_{\mathrm{es}}^n,\\
\rho_{\mathrm{lin}}^n
&\equiv\rho^n-\nabla\cdot\mathbf P_{\mathrm{num}}^n
=\nabla\cdot\mathbf D_{\mathrm{es}}^n.
\end{aligned}
\label{eq:implicit_charge_relation}
\end{equation}
Here $\mathbf P_{\mathrm{num}}$ represents the numerical polarization
associated with the particle correction; $\rho_{\mathrm{lin}}$ is the
corresponding linearized charge density, rather than a new deposition of
the fully corrected particles. Thus $\chi$ represents part of the particle
response and does not add a material dielectric polarization to
$\varepsilon_r$. The same decomposition applies on the grid when both
terms use the same face interpolation and discrete divergence.

The discrete operator uses face-centered effective permittivities; for
example, the coefficient of the $z-$ face of node $(i,j)$ is
\begin{equation}
\varepsilon^{\mathrm{eff}}_{z-}(i,j)
=\left(1+\frac{\chi_{i,j}+\chi_{i-1,j}}{2}\right)
\frac{\varepsilon_{r,i-1,j-1}+\varepsilon_{r,i-1,j}}{2}
\label{eq:epseff}
\end{equation}
and the other three faces are defined analogously. The axis $R=0$ is a
geometrical symmetry boundary, not a metal wall or an electrical ground.
Axisymmetry imposes the electrostatic Neumann condition
\begin{equation}
\left.\frac{\partial\phi}{\partial R}\right|_{R=0}=0,
\qquad E_R(z,0)=-\left.\frac{\partial\phi}{\partial R}\right|_{R=0}=0
\label{eq:axis_electrostatic_bc}
\end{equation}
Metal electrodes and coil conductors, by contrast, enter the discrete system
as Dirichlet boundary values. The potential of an ordinary metal electrode
is imposed by the existing electrode-circuit module. Each coil turn is
assumed to be an equipotential body, so that all conductor nodes of turn $k$
share the single turn potential $\phi_k(t_n)$ reconstructed from the complex
node voltage of the distributed RF network.

The external circuit provides the peak phasor of the node voltage of turn
$k$, $\hat V_k=V_{c,k}-\mathrm{i}V_{s,k}$. Under this equipotential-turn
assumption, the real Dirichlet potential
\begin{equation}
\phi_k(t_n)=V_{c,k}\cos\theta_n+V_{s,k}\sin\theta_n
=\operatorname{Re}\!\left(\hat V_k e^{\mathrm{i}\theta_n}\right)
\label{eq:dirichlet}
\end{equation}
is imposed every PIC step on all conductor nodes of turn $k$ in the Poisson
problem~\eqref{eq:poisson}.

Conductor charge is the coupling variable in established electrostatic
PIC--circuit formulations
\cite{verboncoeur1993simultaneous}. The present capacitive
branch extends the charge-based two-dimensional PIC/MCC coupling used by
Chen et al.~\cite{chen2024gec} from powered electrodes to
separate coil turns. Whereas that earlier formulation closed the electrode
voltage through
an equivalent capacitance, the resolved coil--plasma domain is not reduced
here to a single lumped capacitance. Instead, after the Poisson solve and a
one-layer halo exchange of $\phi$, the discrete circuit charge of each turn
is evaluated from the flux of the Poisson operator. For a face $f$ between
a conductor node of turn $k$ and an external neighbour,
\begin{equation}
\Delta Q_f
=\varepsilon_0\,\varepsilon^{\mathrm{eff}}_f\,
\frac{\phi_k-\phi_{\mathrm{nb}}}{d_f}\,A_f
\label{eq:facecharge}
\end{equation}
where $\varepsilon^{\mathrm{eff}}_f$ is exactly the face
coefficient~\eqref{eq:epseff} of the Poisson operator (including the implicit
susceptibility $\chi$ and the cell permittivities), $d_f$ is the grid spacing
normal to the face, and the true axisymmetric face areas are
\begin{equation}
\begin{aligned}
A_{z\pm}&=2\pi R_j\,\Delta R,\\
A_{R-}&=2\pi\!\left(R_j-\frac{\Delta R}{2}\right)\Delta z,\\
A_{R+}&=2\pi\!\left(R_j+\frac{\Delta R}{2}\right)\Delta z
\end{aligned}
\label{eq:areas}
\end{equation}
The instantaneous charge of turn $k$ is the sum over its external surface,
$Q_k(t_n)=\sum_{f\in\partial\Omega_k}\Delta Q_f$. With $\mathbf n_k$
pointing from the conductor into the surrounding domain,
Eq.~\eqref{eq:implicit_charge_relation} gives
\begin{equation}
\begin{aligned}
Q_k^{\mathrm{free}}&=\sum_{f\in\partial\Omega_k}
\mathbf D_{\mathrm{es},f}\cdot\mathbf n_k\,A_f,\\
Q_k-Q_k^{\mathrm{free}}&=\sum_{f\in\partial\Omega_k}
\mathbf P_{\mathrm{num},f}\cdot\mathbf n_k\,A_f.
\end{aligned}
\label{eq:coil_charge_polarization}
\end{equation}
Here $Q_k^{\mathrm{free}}$ is the discrete estimate of the physical free
charge from the material displacement flux. In the GEC geometry below,
the turns lie in a particle-free compartment outside the dielectric window.
The particle-deposition support does not reach either node used to average
$\chi$ on any coil sampling face, so $\chi_f=0$ there and
$Q_k=Q_k^{\mathrm{free}}$. Plasma and sheath loading still affect this
charge through the total potential obtained from the global Poisson solve.
Using the same face coefficient in the solve and the charge extraction
preserves their discrete Gauss-law consistency. For a sampling surface
with nonzero numerical-polarization flux, however, this consistency alone
does not identify $Q_k$ with the physical conductor charge: the second
line of Eq.~\eqref{eq:coil_charge_polarization} must also be accounted for.
A conductor that collects particles additionally requires the collected
particle current in its circuit balance.

Over one closed RF period the charge harmonics are accumulated as in
Eq.~\eqref{eq:fourier}, giving $\hat Q_k=Q_{c,k}-\mathrm{i}Q_{s,k}$. The
displacement-current phasor follows from the exact fundamental-frequency
derivative,
\begin{equation}
\hat I_{\mathrm{cap},k}=\mathrm{i}\omega\hat Q_k
=\omega Q_{s,k}+\mathrm{i}\omega Q_{c,k}
\label{eq:icap}
\end{equation}
corresponding to the physical current
$I_{\mathrm{cap},k}(t)=\mathrm{d}Q_k/\mathrm{d}t$. No backward difference of
the noisy $Q_k(t_n)$ sequence is formed: extracting the target harmonic
first and then multiplying by the exact factor $\mathrm{i}\omega$ avoids both
noise amplification and the additional phase error of a discrete derivative.
At the end of a cycle, the number of accumulated samples is required to equal
$N_{\mathrm{RF}}$ exactly; otherwise, the circuit update is rejected.

The displacement current~\eqref{eq:icap} already contains the vacuum gaps,
dielectric windows, sheaths and plasma response resolved inside the Poisson
domain. These contributions must therefore not be added again as lumped
capacitors in the circuit. Only genuinely out-of-domain stray capacitances
$C_{\mathrm{stray},k}$ (cables, matching box, \emph{etc.}) enter the network
explicitly through
\begin{equation}
\hat I_{\mathrm{stray},k}=\mathrm{i}\omega C_{\mathrm{stray},k}\hat V_k
\label{eq:stray}
\end{equation}

\subsection{Distributed circuit coupling, power control, and power comparison}
\label{subsec:circuit}

The present implementation uses a single-ended-fed, far-end-grounded
topology: the geometric turn order $1\rightarrow N$ coincides with the
electrical order from the feed to the ground, the shunt branch at node
$\hat V_k$ precedes the series segment carrying $\hat I_k$, and the end of
the last segment is grounded,
\begin{equation}
\hat V_{N+1}=0
\label{eq:ground}
\end{equation}
The winding signs $s_k$ affect only the inductive excitation, not the
electrical connection. The unknown vector of the network is
\begin{equation}
\mathbf x=
\begin{bmatrix}
\hat I_f & \hat I_1 & \cdots & \hat I_N & \hat V_1 & \cdots & \hat V_N
\end{bmatrix}^{\mathrm T}
\label{eq:unknowns}
\end{equation}
containing $2N+1$ complex unknowns (e.g.\ $11$ for a five-turn coil).
This mixed node-voltage/branch-current construction follows the modified
nodal-analysis principle of augmenting node potentials with selected branch
currents~\cite{ho1975modified}. The matrix below is a topology-specific
phasor system for the ordered single-ended coil, rather than a general
netlist-level MNA implementation.

Let $\hat V_g$ denote the Thevenin open-circuit peak phasor, with magnitude
$V_{g,\mathrm{pk}}$ and phase $\varphi_g$. The impedance of the generator plus
the series matching branch is
\begin{equation}
Z_{\mathrm{source}}
=R_g+R_m+\mathrm{i}\left(\omega L_m-\frac{1}{\omega C_m}\right)
\label{eq:zsource}
\end{equation}
with the capacitive term omitted when $C_m\le0$. The per-segment impedance
combines the self-consistently calculated vacuum impedance matrix with the
turn-resolved RF conductor resistance,
\begin{equation}
Z_{\mathrm{turn},kl}
=Z_{\mathrm{vac},kl}+R_{\mathrm{rf},k}\delta_{kl}
\label{eq:zturn_field}
\end{equation}
For a manually supplied total coil resistance distributed uniformly among
the turns, $R_{\mathrm{rf},k}=R_{\mathrm{coil}}/N$. Alternatively, the RF
resistance of each turn is computed from its geometry with a round-wire
skin-effect model. The copper resistivity is corrected for temperature,
$\rho(T)=\rho_{20}\left[1+\alpha_{\mathrm{Cu}}(T-20\,^{\circ}\mathrm C)\right]$,
the skin depth is $\delta=\sqrt{2\rho/(\omega\mu_0)}$, and
\begin{equation}
R_{\mathrm{rf},k}
=\frac{\rho\,(2\pi R_k)}
{\pi\!\left[R_{w,k}^2-\max(R_{w,k}-\delta,\,0)^2\right]}
\label{eq:skin}
\end{equation}
where $R_k$ is the mean radius of the turn and $R_{w,k}$ the equivalent wire
radius inferred from the discrete cross-section.

The feed node obeys the Thevenin equation
\begin{equation}
Z_{\mathrm{source}}\hat I_f+\hat V_1=\hat V_g
\label{eq:thevenin}
\end{equation}
Defining the node input currents $\hat I_{\mathrm{in},1}=\hat I_f$ and
$\hat I_{\mathrm{in},k}=\hat I_{k-1}$ ($k>1$), Kirchhoff's current law at
node $k$ reads
\begin{equation}
\hat I_{\mathrm{in},k}-\hat I_k
-\mathrm{i}\omega C_{\mathrm{stray},k}\hat V_k
=\hat I_{\mathrm{cap},k}
\label{eq:kcl}
\end{equation}
Here positive $\hat I_{\mathrm{cap},k}$ is directed from node $k$ into the
spatial Poisson domain. This convention makes it the shunt current drawn by
the resolved capacitive plasma--dielectric load.
Kirchhoff's voltage law along segment $k$ reads
\begin{equation}
\hat V_k-\hat V_{k+1}
-\sum_{l=1}^{N}Z_{\mathrm{turn},kl}\hat I_l
=\hat V_{p,k}^{\mathrm{ind}},
\qquad \hat V_{N+1}=0
\label{eq:kvl}
\end{equation}
Here $\hat V_{p,k}^{\mathrm{ind}}$ [Eq.~\eqref{eq:vpind}] is the controlled
voltage source of the inductive channel and $\hat I_{\mathrm{cap},k}$
[Eq.~\eqref{eq:icap}] is the controlled current source of the capacitive
channel; both are measured directly by the PIC model during the completed RF
cycle.

With $\mathbf e_1=[1,0,\ldots,0]^{\mathrm T}$, the diagonal stray-admittance
matrix
$\mathbf Y_s=\operatorname{diag}(\mathrm{i}\omega C_{\mathrm{stray},1},\ldots,
\mathrm{i}\omega C_{\mathrm{stray},N})$, the current-difference matrix
$B_{kl}=-\delta_{kl}+\delta_{k-1,l}\,(k>1)$ and the voltage-difference matrix
$D_{kl}=\delta_{kl}-\delta_{k+1,l}\,(k<N)$, the full system takes the block
form
\begin{equation}
\begin{bmatrix}
Z_{\mathrm{source}} & \mathbf 0^{\mathrm T} & \mathbf e_1^{\mathrm T}\\
\mathbf e_1 & \mathbf B & -\mathbf Y_s\\
\mathbf 0 & -\mathbf Z_{\mathrm{turn}} & \mathbf D
\end{bmatrix}
\begin{bmatrix}
\hat I_f\\[0.5ex]
\hat{\mathbf I}\\[0.5ex]
\hat{\mathbf V}
\end{bmatrix}
=
\begin{bmatrix}
\hat V_g\\[0.5ex]
\hat{\mathbf I}_{\mathrm{cap}}\\[0.5ex]
\hat{\mathbf V}_{p}
\end{bmatrix}
\label{eq:blockmatrix}
\end{equation}
This form separates three physical ingredients: the external linear elements
reside in the coefficient matrix, the nonlinear load responses measured by
the PIC model during the completed cycle enter the right-hand side, and the
coil currents and node voltages of the next cycle are the unknowns.

The measured displacement current is used directly as a controlled source
rather than being converted into a per-turn admittance
$Y_{\mathrm{cap},k}=\hat I_{\mathrm{cap},k}/\hat V_k$ in the matrix diagonal.
This choice avoids the numerical singularity and PIC-noise amplification of
the ratio $I/V$ when a node voltage crosses zero during startup, retains the
explicit one-cycle time level of the nonlinear sheath/plasma response, and
leaves the KCL residual directly checkable.

Let $\mathbf x^n=(\hat I_f^n,\hat{\mathbf I}^n,\hat{\mathbf V}^n)$ be the
circuit state actually applied during RF cycle $n$: its phasors are held
fixed for the whole cycle, while the real fields and boundary values are
reconstructed every PIC step according to the phase. Thus,
$\mathbf x^n\equiv\mathbf x_{\mathrm{applied}}^{\,n}$; the cycle superscript
distinguishes this outer update from the PIC steps within the period.
At the end of cycle $n$, the load measurements
\begin{equation}
\hat{\mathbf V}_p^{\,n}
=\mathcal F_{\mathrm{ind}}(\text{PIC state};\mathbf x^n),
\qquad
\hat{\mathbf I}_{\mathrm{cap}}^{\,n}
=\mathcal F_{\mathrm{cap}}(\text{PIC state};\mathbf x^n)
\label{eq:loadmeasurement}
\end{equation}
are finalized together with the completed-cycle power diagnostics. These
load responses are then held fixed throughout the circuit update described
below.

An optional closed-loop controller adjusts the Thevenin peak voltage so that
the active power at the coil terminal approaches a target value. The feedback
signal is the terminal active power of the completed cycle computed by the
circuit itself,
$P_{\mathrm{port}}^{\,n}=\tfrac12\operatorname{Re}(\hat V_1^{\,n}\hat I_f^{\,n*})$;
the independently accumulated field-side powers are used only as diagnostics
and never enter the controller. The controller is evaluated before the
next target network is solved. The measurement is filtered,
\begin{equation}
P_f^{\,n}=\gamma P_{\mathrm{port}}^{\,n}+(1-\gamma)P_f^{\,n-1}
\label{eq:pfilter}
\end{equation}
and its per-cycle trend is extrapolated to compensate for the delayed plasma
response,
\begin{equation}
P_c^{\,n}
=\max\!\left[P_f^{\,n}
+N_{\mathrm{trend}}\left(P_f^{\,n}-P_f^{\,n-1}\right),\;
\epsilon_P\right]
\label{eq:ptrend}
\end{equation}
The desired and relaxed source voltages are
\begin{equation}
\begin{aligned}
V_{g,\mathrm{desired}}
&=V_{g,\mathrm{old}}\sqrt{\frac{P_{\mathrm{target}}}{P_c^{\,n}}},\\
V_{g,\mathrm{new}}
&=V_{g,\mathrm{old}}
+\alpha_V\left(V_{g,\mathrm{desired}}-V_{g,\mathrm{old}}\right)
\end{aligned}
\label{eq:pcontrol}
\end{equation}
followed by one symmetric limiter
$(1\pm f_{\mathrm{step}})V_{g,\mathrm{old}}$ and by the absolute voltage
bounds $[V_{\min},V_{\max}]$. The controller changes only the generator
amplitude, never its phase. Here $V_{g,\mathrm{old}}$ is the amplitude at
cycle $n$, and the limited $V_{g,\mathrm{new}}$ defines the source phasor
$\hat V_g^{\,n+1}$ supplied to the next network solve. With power control
disabled, the prescribed source voltage is used instead.

The target state is next obtained from the updated source voltage and the
two load responses measured in cycle $n$,
\begin{equation}
\mathbf x_{\mathrm{target}}^{n+1}
=\mathcal C\!\left(\hat{\mathbf V}_p^{\,n},
\hat{\mathbf I}_{\mathrm{cap}}^{\,n},\hat V_g^{\,n+1}\right)
\label{eq:target}
\end{equation}
To damp cycle-to-cycle oscillations caused by PIC noise and by the nonlinear
sheath load, the feed current, all turn currents and all turn voltages are
then relaxed with one common factor $\alpha$,
\begin{equation}
\mathbf x_{\mathrm{applied}}^{\,n+1}
=(1-\alpha)\,\mathbf x_{\mathrm{applied}}^{\,n}
+\alpha\,\mathbf x_{\mathrm{target}}^{\,n+1},
\qquad 0<\alpha\le1
\label{eq:relaxation}
\end{equation}
A single factor preserves the relative update scales of voltages and currents
within the same network solution. The resulting $\mathbf x^{n+1}$ is
broadcast and used to reconstruct the boundary voltages and azimuthal field
for cycle $n+1$. The load response is not remeasured during this network
solve. The sequence is therefore completed-cycle measurement, source-voltage
update, target network solve, state relaxation, and next-cycle application,
with one explicit RF-period delay in the block-wise partitioned fixed-point
iteration.

To assess the coupled state independently of the controller, the field-side
power is compared with the circuit-side port power, and the KCL/KVL residuals
are monitored. The inductive power deposited in the plasma is accumulated
every PIC step independently of the circuit,
\begin{equation}
P_{\mathrm{ind}}
=\frac{1}{N_{\mathrm{RF}}}\sum_{n=1}^{N_{\mathrm{RF}}}
\int_{\Omega}J_\theta(t_n)\,E_\theta(t_n)\,\mathrm{d}V
\label{eq:pind}
\end{equation}
Since $I_{\mathrm{cap}}\,\mathrm{d}t=\mathrm{d}Q$, the capacitive power
transferred from the coil turns to the resolved electrostatic domain is
evaluated as the closed-cycle trapezoidal integral of $V\,\mathrm{d}Q$,
\begin{equation}
W_k=\sum_{n=1}^{N_{\mathrm{RF}}}
\frac{V_{k,n}+V_{k,n-1}}{2}\left(Q_{k,n}-Q_{k,n-1}\right),
\qquad
P_{\mathrm{cap}}=f\sum_{k=1}^{N}W_k
\label{eq:pcap}
\end{equation}
where the last interval is closed periodically between the final and the
first sample of the cycle, consistent with the harmonic assumption. This is
a discrete realization of
$P_{\mathrm{cap},k}=T_{\mathrm{RF}}^{-1}\oint V_k\,\mathrm{d}Q_k$.
Active power is associated with the current component in phase with the
voltage, or equivalently with the charge component in quadrature with it.
For example, with $V=V_0\cos\omega t$ and
$Q=Q_c\cos\omega t+Q_s\sin\omega t$, the adopted convention gives
$\langle V\,\mathrm{d}Q/\mathrm{d}t\rangle_{\mathrm{RF}}
=\omega V_0Q_s/2$. A lossless linear capacitor has $Q_s=0$ in this voltage
reference and hence zero cycle-averaged active power. As a cross-check, the
phasor power
\begin{equation}
P_{\mathrm{cap}}^{\mathrm{ph}}
=\frac12\operatorname{Re}\sum_{k=1}^{N}
\hat V_k\,\hat I_{\mathrm{cap},k}^{*}
\label{eq:pcap_phasor}
\end{equation}
is evaluated alongside. For uniform closed sampling, a purely
fundamental-frequency imposed voltage, and the same charge samples,
Eqs.~\eqref{eq:pcap} and~\eqref{eq:pcap_phasor} obey
\begin{equation}
P_{\mathrm{cap}}
=\frac{\sin(2\pi/N_{\mathrm{RF}})}{2\pi/N_{\mathrm{RF}}}
P_{\mathrm{cap}}^{\mathrm{ph}}.
\label{eq:pcap_quadrature}
\end{equation}
This comparison checks the harmonic phase convention and cycle quadrature;
both estimates use the same conductor-charge samples. For a lossless
linear network $\hat{\mathbf Q}=\mathbf C_M\hat{\mathbf V}$ with a real
symmetric capacitance matrix, both yield zero total active power and the
reactive power
$Q=-\tfrac12\omega\sum_{kl}C_{M,kl}\hat V_k\hat V_l^{*}$, which serves as the
analytical benchmark of the implementation.

The relation between this port work and plasma heating follows from the
continuum electroquasistatic energy balance. Consider fixed, lossless,
nondispersive dielectrics, coil conductors insulated from particle
collection, and all other electrical boundaries grounded, as in the GEC
application below. Charge continuity and Gauss's law then give
\begin{equation}
\sum_{k=1}^{N}V_k(t)\frac{\mathrm{d}Q_k}{\mathrm{d}t}
=\frac{\mathrm{d}W_{\mathrm{es}}}{\mathrm{d}t}
+\int_{\Omega_p}(J_zE_z+J_RE_R)\,\mathrm{d}V,
\label{eq:es_power_balance}
\end{equation}
where $\Omega_p$ is the plasma region and $J_z,J_R$ include the currents
of all charged-particle species. The physical electrostatic energy is
$W_{\mathrm{es}}=\tfrac12\int_{\Omega}\varepsilon_0\varepsilon_r
(E_z^2+E_R^2)\,\mathrm{d}V$, including storage in the plasma, sheaths,
vacuum gaps, and dielectric. The numerical susceptibility $\chi$ in
Eq.~\eqref{eq:poisson} represents the implicit particle response and is not
an additional material contribution to this stored energy.

At periodic steady state,
$\langle\mathrm{d}W_{\mathrm{es}}/\mathrm{d}t\rangle_{\mathrm{RF}}
=[W_{\mathrm{es}}(t+T_{\mathrm{RF}})-W_{\mathrm{es}}(t)]/T_{\mathrm{RF}}=0$.
The stored energy may oscillate within a period; its net change over the
period vanishes. Equation~\eqref{eq:es_power_balance} therefore gives
\begin{equation}
P_{\mathrm{cap}}
=\left\langle\int_{\Omega_p}(J_zE_z+J_RE_R)\,\mathrm{d}V
\right\rangle_{\mathrm{RF}}.
\label{eq:pcap_heating}
\end{equation}
This continuum equality identifies the port power with net capacitive
plasma heating under the stated conditions; the discrete diagnostic
approximates it subject to PIC discretization and cycle-quadrature errors.
For a statistically stationary PIC discharge, the vanishing storage term is
understood as an average over the steady-state window. Heating here denotes
net electrostatic work on electrons and ions. Collisional energy transfer
and particle energy carried to or from surfaces, including secondary
emission, enter the particle energy balance and need not vanish at steady
state. Dielectric dissipation, leakage, or another biased electrode would
require additional terms in Eq.~\eqref{eq:es_power_balance}. Copper loss is
accounted for separately in the coil circuit. The port-work diagnostic does
not by itself resolve electron-only heating or the subsequent partition of
particle energy among collisions and wall exchange.

The circuit-side terminal active power of the completed cycle is
\begin{equation}
P_{\mathrm{port}}
=\frac12\operatorname{Re}
\!\left(\hat V_1^{\,n}\hat I_f^{\,n*}\right)
\label{eq:pport}
\end{equation}
and the copper loss of the same cycle is
$P_{\mathrm{Cu}}=\tfrac12\sum_k R_{\mathrm{rf},k}|\hat I_k^{\,n}|^2$.
The power balance diagnostic
\begin{equation}
P_{\mathrm{ind}}+P_{\mathrm{cap}}+P_{\mathrm{Cu}}
\;\simeq\;P_{\mathrm{port}}
\label{eq:pbalance}
\end{equation}
compares two independently accumulated quantities (field-side integrals
versus circuit phasors) and is monitored every cycle. All terms refer to the
completed cycle. Before fixed-point convergence, the applied circuit state
need not satisfy the network equations for the plasma response measured
over that cycle. Moreover, Eq.~\eqref{eq:pcap} imposes periodic charge closure
and does not measure secular charging, so Eq.~\eqref{eq:pbalance} is not a
complete transient energy balance. Equality is expected after the cycle
fixed point and the plasma statistical steady state have been reached.
For the final converged state, the relative power imbalance is evaluated
with the terminal power as the reference,
\begin{equation}
\epsilon_{\mathrm{bal}}
=\frac{\left|P_{\mathrm{port}}
-\left(P_{\mathrm{ind}}+P_{\mathrm{cap}}+P_{\mathrm{Cu}}\right)\right|}
{\left|P_{\mathrm{port}}\right|}
<1\%
\label{eq:power_imbalance}
\end{equation}

The per-node KCL and per-segment KVL residuals are evaluated after the
update, using the state prepared for cycle $n+1$ and the load responses
measured in cycle $n$. With these cycle indices suppressed, they are
\begin{equation}
r_{I,k}
=\hat I_{\mathrm{in},k}-\hat I_k-\hat I_{\mathrm{cap},k}
-\mathrm{i}\omega C_{\mathrm{stray},k}\hat V_k
\label{eq:kclres}
\end{equation}
\begin{equation}
r_{V,k}
=\hat V_k-\hat V_{k+1}-\sum_{l}Z_{\mathrm{turn},kl}\hat I_l
-\hat V_{p,k}^{\mathrm{ind}},
\qquad
r_{V,0}=\hat V_g-Z_{\mathrm{source}}\hat I_f-\hat V_1
\label{eq:kvlres}
\end{equation}
The residuals are normalized by the corresponding current and voltage
scales, and their maxima are recorded every cycle. In these expressions,
the circuit currents, node voltages and $\hat V_g$ carry index $n+1$,
whereas $\hat I_{\mathrm{cap},k}$ and $\hat V_{p,k}^{\mathrm{ind}}$ carry
index $n$. For $\alpha=1$, the applied state equals the target solution,
so these residuals check the network assembly and linear solve for the
frozen loads. For $\alpha<1$, they also include the effect of state
relaxation and monitor the approach to the coupled fixed point. Small
residuals for a frozen load alone do not establish that the plasma response
has become stationary.

Here steady state requires both a settled circuit operating point and a
statistically stationary plasma response. In the noise-free periodic limit,
successive circuit states and load phasors coincide, and the controlled
source amplitude no longer changes. For PIC/MCC, statistical stationarity
means that the cycle-averaged plasma density and mean electron energy,
circuit phasors, and power partition have no systematic drift over the
sampling window, while statistical fluctuations may remain. Reaching
$P_{\mathrm{target}}$ alone does not establish this joint condition.
Likewise, Eq.~\eqref{eq:power_imbalance} checks energy consistency of the
final state; it is not by itself a stopping criterion for the coupled
iteration.

\subsection{Overall algorithm workflow}
\label{subsec:workflow}

\begin{figure*}[!htbp]
\centering
\includegraphics[width=\textwidth]{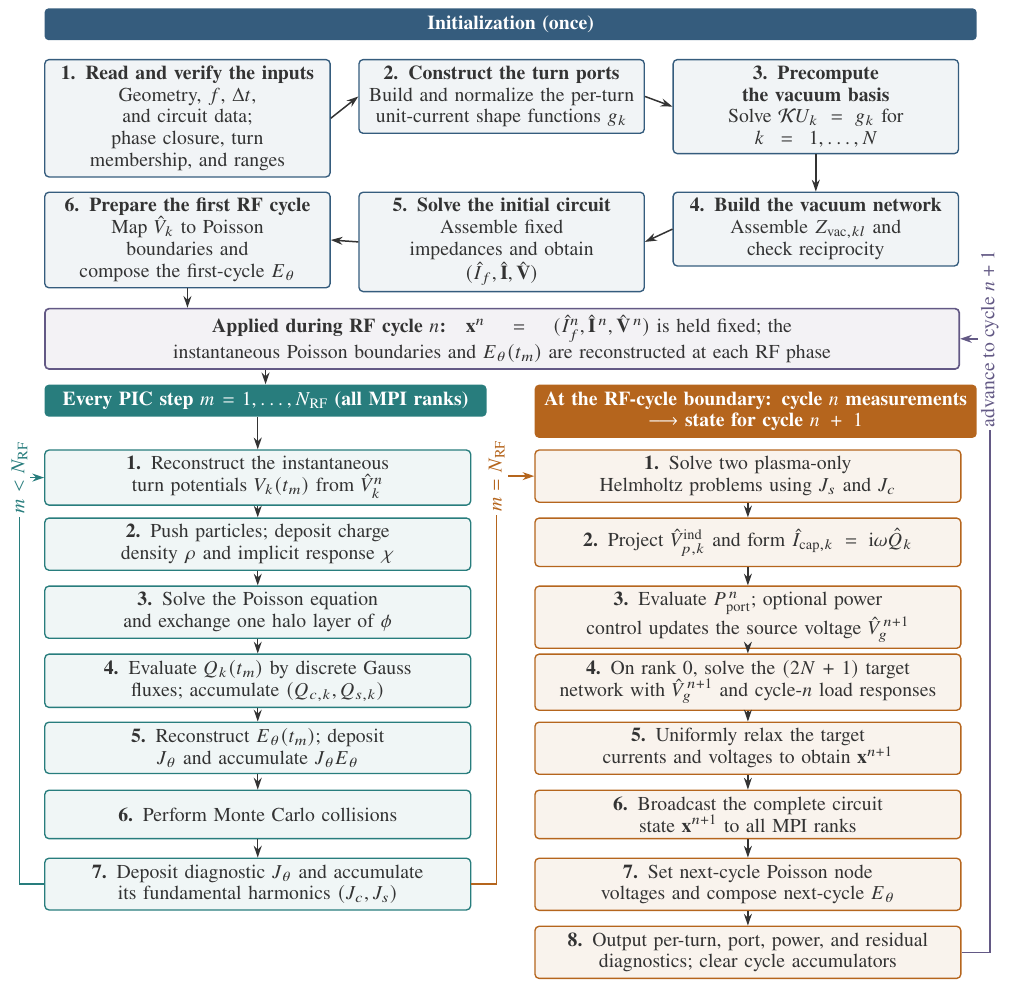}
\caption{Complete workflow of the bidirectionally coupled phasor--PIC
algorithm. The initialization is executed once. During RF cycle $n$, the
applied circuit phasors $\mathbf x^n$ remain fixed while
$N_{\mathrm{RF}}$ PIC steps accumulate the capacitive and inductive plasma
responses. At the cycle boundary, the completed-cycle port power first
updates the source voltage when power control is enabled. The target circuit
state is then solved on rank~0 using this source and the cycle-$n$ load
measurements. Uniform relaxation gives $\mathbf x^{n+1}$, which is broadcast
and used to reconstruct the
Poisson boundary values and azimuthal field for the next cycle. The outer
feedback arrow therefore represents the explicit one-RF-period delay in the
block-wise fixed-point iteration.}
\label{fig:algorithm_workflow}
\end{figure*}

Figure~\ref{fig:algorithm_workflow} summarizes the three-level execution
sequence and the one-RF-period-delayed fixed-point feedback. Initialization
validates the configuration, constructs the normalized turn sources $g_k$,
solves the unit fields $U_k$, assembles $Z_{\mathrm{vac}}$ and the fixed
network, and generates the first-cycle field and circuit state.

During RF cycle $n$, the circuit phasors $\mathbf x^n$ remain fixed for
$N_{\mathrm{RF}}$ PIC/MCC steps. Their instantaneous turn potentials and
azimuthal field drive the particle and Poisson updates, while the turn
charges, plasma-current harmonics and power diagnostics are accumulated. At
the cycle boundary, two plasma-only Helmholtz solves provide the inductive
back-emfs, and the charge harmonics provide the capacitive currents. On
rank~0, the optional power controller first uses $P_{\mathrm{port}}^{\,n}$
to update $\hat V_g^{\,n+1}$. The $(2N+1)$ network is then solved with this
source voltage and the cycle-$n$ load responses. Uniform relaxation gives
$\mathbf x^{n+1}$, which is broadcast and used to construct the next-cycle
fields. The completed-cycle powers are retained for their same-cycle
comparison in Eq.~\eqref{eq:pbalance}. This update order produces the explicit
one-cycle delay shown by the outer feedback arrow in
Fig.~\ref{fig:algorithm_workflow}.

\FloatBarrier

\section{Results}
\label{sec:results}

Verification progresses from the inductive components to the joint
field--circuit closure. Hierarchical benchmarks help distinguish the
components and couplings tested by each verification
problem~\cite{eckert2025benchmark}. The first
two component tests compare the diagonal and off-diagonal inductive paths
with analytical solenoid references and verify their transfer to the phasor
circuit. The underlying two-dimensional electrostatic PIC--circuit coupling
was introduced and verified previously~\cite{chen2024gec}. The third test
examines its integration
with the inductive channel in one distributed network. A five-node
manufactured MQS--EQS problem activates both field responses and compares
all eleven complex unknowns and the period-delayed iteration with independent
references. With particles absent, this comparison isolates the linear
field--circuit response and quantifies its finite-grid errors.
Section~\ref{subsec:application} then applies the complete bidirectional
algorithm to an argon plasma in a two-dimensional GEC reference cell, where
the evidence comprises the spatial plasma response, an experimental density-
profile comparison, the loaded port impedance, and independent field--circuit
power closure. Table~\ref{tab:coverage} summarizes the field operator,
plasma response, exercised feedback path, and comparison basis of each
configuration.

\begin{table*}[!tbp]
\centering
\small
\caption{Coverage of the verification configurations and the GEC
application. Tests 1--3 are particle-free and use the magnetoquasistatic
(MQS) inductive operator, i.e.\ Eq.~\eqref{eq:helmholtz_phasor} without the
displacement-current term. The GEC application retains the full operator and
exercises both feedback channels with the self-consistent plasma response,
including the nonzero plasma back-emf.}
\label{tab:coverage}
\begin{tabularx}{\textwidth}{@{}
>{\raggedright\arraybackslash}p{2.2cm}
>{\hsize=0.85\hsize\raggedright\arraybackslash}X
>{\hsize=1.10\hsize\raggedright\arraybackslash}X
>{\hsize=1.05\hsize\raggedright\arraybackslash}X@{}}
\toprule
 & Tests 1--2 & Test 3 & GEC application\\
\midrule
inductive operator
 & MQS vacuum operator
 & MQS operator
 & full $\mathcal L$, displacement-current term retained\\
\addlinespace[3pt]
electrostatic solve
 & ---
 & Laplace, manufactured boundaries
 & Poisson with implicit susceptibility\\
\addlinespace[3pt]
plasma response
 & none
 & none
 & self-consistent PIC/MCC plasma\\
\addlinespace[3pt]
feedback exercised
 & inductive self/mutual path
 & joint inductive--capacitive closure; period-delayed update
 & joint bidirectional coupling; nonzero plasma back-emf\\
\addlinespace[3pt]
comparison basis
 & analytical solenoid references
 & continuous and semidiscrete references; four-grid refinement
 & measured density profile; internal power closure\\
\bottomrule
\end{tabularx}
\end{table*}

\FloatBarrier

\subsection{Test 1: vacuum single ideal solenoid}
\label{subsec:verif_solenoid}

\begin{figure*}[!htbp]
\centering
\includegraphics[width=\textwidth]{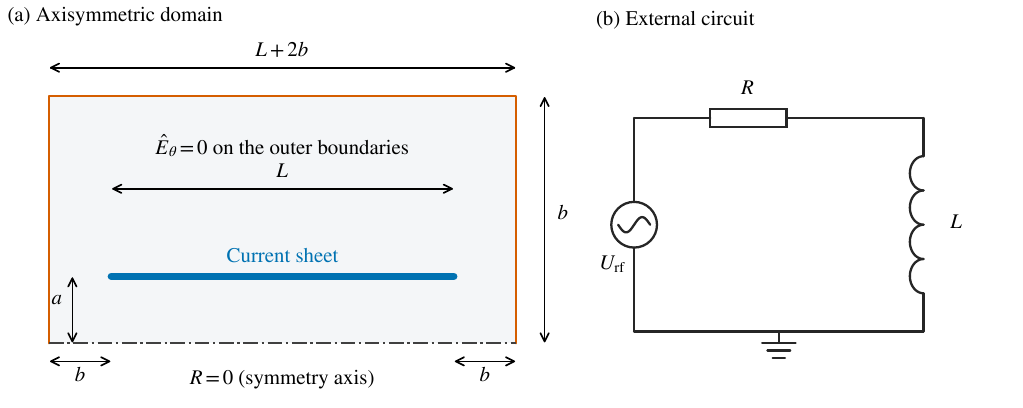}
\caption{Single-solenoid verification model (not to scale).
(a)~Axisymmetric vacuum domain containing an azimuthal current sheet of
radius $a$ and length $L$, with an axial padding distance $b$ at each end.
The $R=0$ edge is the symmetry axis. The sheet is an internal current source,
and the remaining edges are finite homogeneous Dirichlet boundaries for
$\hat E_\theta$. (b)~Series external circuit driven by
$U_{\mathrm{rf}}$, with resistance $R$ and the inductance obtained from the
field-derived vacuum impedance. The symbols $R$ and $L$ therefore denote
radial coordinate and sheet length in panel~(a), but resistance and
inductance in panel~(b).}
\label{fig:S1model}
\end{figure*}

This test verifies the complete single-port inductive path in vacuum, from
the normalized coil source and Helmholtz field solve to the field-derived
vacuum impedance and the series-phasor circuit update. As illustrated in
Fig.~\ref{fig:S1model}(a), the ideal solenoid is represented in the
axisymmetric $R$--$z$ plane by a thin cylindrical current sheet of radius
$a$ and axial length $L$. The sheet carries the total ampere-turn current
$I_p=NI_w$, where $N$ is the number of turns and $I_w$ is the wire current.
Its uniform azimuthal surface-current density is therefore
$K_\theta=I_p/L$. For the positive winding direction used here,
$K_\theta>0$ and the magnetic field inside the sheet points in the $+z$
direction.

The current sheet is an internal volume-source region of the Helmholtz
problem, not a field boundary. The edge $R=0$ is the symmetry axis rather
than a metal or grounded wall. Regularity and the three finite homogeneous
Dirichlet truncation boundaries give
\begin{equation}
\begin{gathered}
\hat E_\theta(z,0)=0,\\
\hat E_\theta(0,R)=\hat E_\theta(L+2b,R)=0,\\
\hat E_\theta(z,b)=0
\end{gathered}
\label{eq:S1bc}
\end{equation}
Figure~\ref{fig:S1model}(b) shows the associated external circuit. A
sinusoidal source $U_{\mathrm{rf}}$ drives the series combination of a
resistance $R$ and the field-derived coil inductance. With no plasma current,
the circuit is closed by
\begin{equation}
\hat I=\frac{\hat U_{\mathrm{rf}}}
{R+\mathrm{i}\omega L_{\mathrm{num}}},\qquad
\hat U_L=\mathrm{i}\omega L_{\mathrm{num}}\hat I,\qquad
L_{\mathrm{num}}=\frac{\operatorname{Im}Z_{\mathrm{vac}}}{\omega}
\label{eq:S1circuit}
\end{equation}
Here $L$ denotes the sheet length in Fig.~\ref{fig:S1model}(a), whereas the
inductance in the circuit is reported as $L_{\mathrm{num}}$ or
$L_{\mathrm{ana}}$ below.

The complete base-case conditions are summarized in
Table~\ref{tab:S1conditions}. The table separates the physical-model,
discretization, and circuit settings; the boundary definitions are those of
Eq.~\eqref{eq:S1bc}.

\begin{table*}[!tbp]
\centering
\caption{Base-case model and numerical conditions for the single-solenoid
test.}
\label{tab:S1conditions}
\begin{tabular}{
>{\raggedright\arraybackslash}p{0.29\textwidth}
>{\raggedright\arraybackslash}p{0.63\textwidth}}
\toprule
item & specification\\
\midrule
Current-sheet geometry
& $a=10$~mm, $L=1.0$~m, $b=0.12$~m; nominal sheet interval
$0.12\le z\le1.12$~m\\
Vacuum domain and mesh
& $0\le z\le1.24$~m, $0\le R\le0.12$~m;
$\Delta z=\Delta R=h=1$~mm;
$N_z\times N_R=1241\times121$\\
Source discretization
& two-cell radial width; 1000 active source nodes centred at
$z=0.121$--$1.120$~m;
$\sum g\,\Delta z\,\Delta R=1$\\
Field model and boundaries
& magnetoquasistatic vacuum model without the displacement-current term;
regularity and finite homogeneous Dirichlet conditions given by
Eq.~\eqref{eq:S1bc}\\
RF sampling
& $f=13.56$~MHz; $N_{\mathrm{RF}}=400$ steps per period; peak-phasor
convention $\exp(+\mathrm{i}\omega t)$\\
Series circuit
& $\hat U_{\mathrm{rf}}=1\angle0^\circ$~V;
$R=0.1~\Omega$; zero coil resistance and zero matching elements;
$Z_{\mathrm{vac}}$ obtained from the field projection\\
Excluded physics and controls
& no particles, plasma current, dielectric, or metallic chamber;
capacitive coupling, plasma loading, and power control disabled;
current under-relaxation factor $\alpha=1$\\
\bottomrule
\end{tabular}
\end{table*}

For the corresponding infinitely long ideal solenoid, the analytical field
and vector potential per ampere-turn are
\begin{equation}
\frac{B_z}{I_p}=
\begin{cases}
\mu_0/L, & R<a,\\
0, & R>a,
\end{cases}
\qquad
\frac{A_\theta}{I_p}=
\begin{cases}
\mu_0R/(2L), & R<a,\\
\mu_0a^2/(2LR), & R>a
\end{cases}
\label{eq:sol_field}
\end{equation}
Under the $\exp(+\mathrm{i}\omega t)$ convention,
$\hat E_\theta=-\mathrm{i}\omega A_\theta$. The stored component is therefore
compared with $-\omega A_\theta/I_p$ in the profiles below. The analytical
inductance, current, and coil-port voltage are
\begin{equation}
L_{\mathrm{ana}}=\mu_0\frac{\pi a^2}{L},\qquad
\hat I_{\mathrm{ana}}=
\frac{\hat U_{\mathrm{rf}}}{R+\mathrm{i}\omega L_{\mathrm{ana}}},
\qquad
\hat U_{L,\mathrm{ana}}=\mathrm{i}\omega L_{\mathrm{ana}}
\hat I_{\mathrm{ana}}
\label{eq:sol_L}
\end{equation}
For comparison along the axis of the finite sheet, the analytical reference
is
\begin{equation}
\begin{aligned}
\frac{B_z(0,z)}{I_p}
&=\frac{\mu_0}{2L}
\left[
\frac{z-z_1}{\sqrt{a^2+(z-z_1)^2}}
-
\frac{z-z_2}{\sqrt{a^2+(z-z_2)^2}}
\right],\\
z_1&=b,\qquad z_2=b+L
\end{aligned}
\label{eq:S1finiteBz}
\end{equation}

\begin{table}[!htbp]
\centering
\small
\setlength{\tabcolsep}{3pt}
\caption{Analytical and model results for the base single-solenoid test.
Current and voltage phasors are peak values in polar form under the
$\exp(+\mathrm{i}\omega t)$ convention. The axial magnetic field is
evaluated at the sheet centre,
$z_{\mathrm{c}}=(z_1+z_2)/2$.}
\label{tab:S1results}
\begin{tabular}{lcc}
\toprule
quantity & analytical solution & model result\\
\midrule
$\operatorname{Im}Z_{\mathrm{vac}}$ [$\Omega$]
& 0.0336356 & 0.0331037\\
$L$ [nH]
& 0.394784 & 0.388541\\
$\hat I$ [A]
& $9.4782\angle-18.59^\circ$ & $9.4934\angle-18.32^\circ$\\
$\hat U_L$ [V]
& $0.3188\angle71.41^\circ$ & $0.3143\angle71.68^\circ$\\
$B_z(0,z_{\mathrm{c}})/I_p$ [$\mu\mathrm{T\,A^{-1}}$]
& 1.25639 & 1.24624\\
\bottomrule
\end{tabular}
\end{table}

\begin{figure*}[!htbp]
\centering
\includegraphics[width=\textwidth]{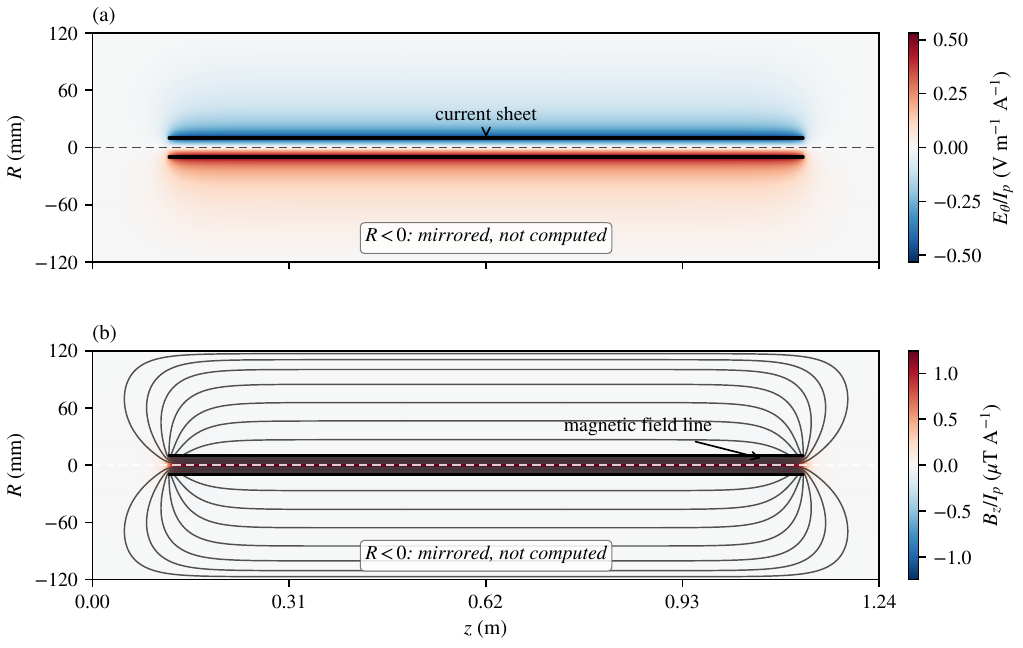}
\caption{Base-case two-dimensional unit fields.
(a)~Azimuthal electric field $E_\theta/I_p$ and
(b)~axial magnetic field $B_z/I_p$ with magnetic-field lines. The
$0\le R\le b=120$~mm half is taken from the numerical solution; the
$-120\le R<0$~mm half is its axisymmetric mirror and was not computed
separately.}
\label{fig:S1maps}
\end{figure*}

Figure~\ref{fig:S1maps} shows the two-dimensional distributions of
$E_\theta/I_p$ and $B_z/I_p$ over $0\le z\le1.24$~m and
$-120\le R\le120$~mm. The numerical model is solved only in the upper
half-plane, $0\le R\le b=120$~mm, and the field over this non-negative
interval is taken directly from the numerical solution. The interval
$-120\le R<0$~mm is generated solely for visualization through the axisymmetric
parities
$E_\theta(-R,z)=-E_\theta(R,z)$ and
$B_z(-R,z)=B_z(R,z)$, and is not an independently computed domain.
The field maps reproduce the expected antisymmetric azimuthal electric field,
the approximately uniform axial magnetic field inside the long sheet, and
the return magnetic field near both ends.

The scalar analytical--model comparison is collected in
Table~\ref{tab:S1results}. The base field projection gives a purely reactive
vacuum impedance. Its field-derived inductance is $1.58$\% lower than the
infinite-solenoid value. Inserting this impedance into
Eq.~\eqref{eq:S1circuit} gives a current-amplitude difference of $0.16$\%, a
coil-voltage-amplitude difference of $1.4$\%, and a phase difference of
$0.27^\circ$ for both quantities.

\begin{table}[!htbp]
\centering
\small
\setlength{\tabcolsep}{3pt}
\caption{Sensitivity of the field-derived inductance to the finite-domain
parameters in the single-solenoid test. The base case is listed first.
The boundary distance $b$ sets both the radial truncation position and the
axial padding on each side of the current sheet.}
\label{tab:S1cases}
\begin{tabular}{lccccc}
\toprule
case & $h$ [mm] & $L$ [m] & $b$ [m] &
$L_{\mathrm{num}}$ [nH] & $\epsilon_L$ [\%]\\
\midrule
base          & 1.0 & 1.0 & 0.12 & 0.388541 & $-1.58$\\
boundary scan & 1.0 & 1.0 & 0.08 & 0.385302 & $-2.40$\\
boundary scan & 1.0 & 1.0 & 0.16 & 0.389653 & $-1.30$\\
grid scan     & 2.0 & 1.0 & 0.12 & 0.387111 & $-1.94$\\
grid scan     & 0.5 & 1.0 & 0.12 & 0.388912 & $-1.49$\\
length scan   & 1.0 & 0.5 & 0.12 & 0.771114 & $-2.34$\\
length scan   & 1.0 & 2.0 & 0.12 & 0.195018 & $-1.20$\\
\bottomrule
\end{tabular}
\end{table}
\FloatBarrier

Table~\ref{tab:S1cases} summarizes the dependence of the field-derived
inductance on the principal finite-domain parameters. All cases use the same
magnetoquasistatic model; each auxiliary pair varies the boundary padding
$b$, mesh spacing $h$, or sheet length $L$ relative to the base case. The
reported error is
$\epsilon_L=(L_{\mathrm{num}}-L_{\mathrm{ana}})/L_{\mathrm{ana}}$, where
$L_{\mathrm{ana}}$ is given by Eq.~\eqref{eq:sol_L}.

All seven cases satisfy $|\epsilon_L|<3$\%. Moving the finite homogeneous
Dirichlet boundaries outward from $b=0.08$ to $0.16$~m reduces the error
magnitude from $2.40$\% to $1.30$\%. By comparison, refining the mesh from
$h=1.0$ to $0.5$~mm at $b=0.12$~m changes the error magnitude by only
$0.09$ percentage points. These trends indicate that the sub-$3$\%
inductance discrepancy is governed primarily by the finite-boundary
truncation rather than by spatial discretization over the tested range. The
length scan also shows a finite-sheet-length sensitivity, so the three
parameter scans are not interpreted as an additive decomposition of
independent errors.

\begin{figure*}[!htbp]
\centering
\includegraphics[width=\textwidth]{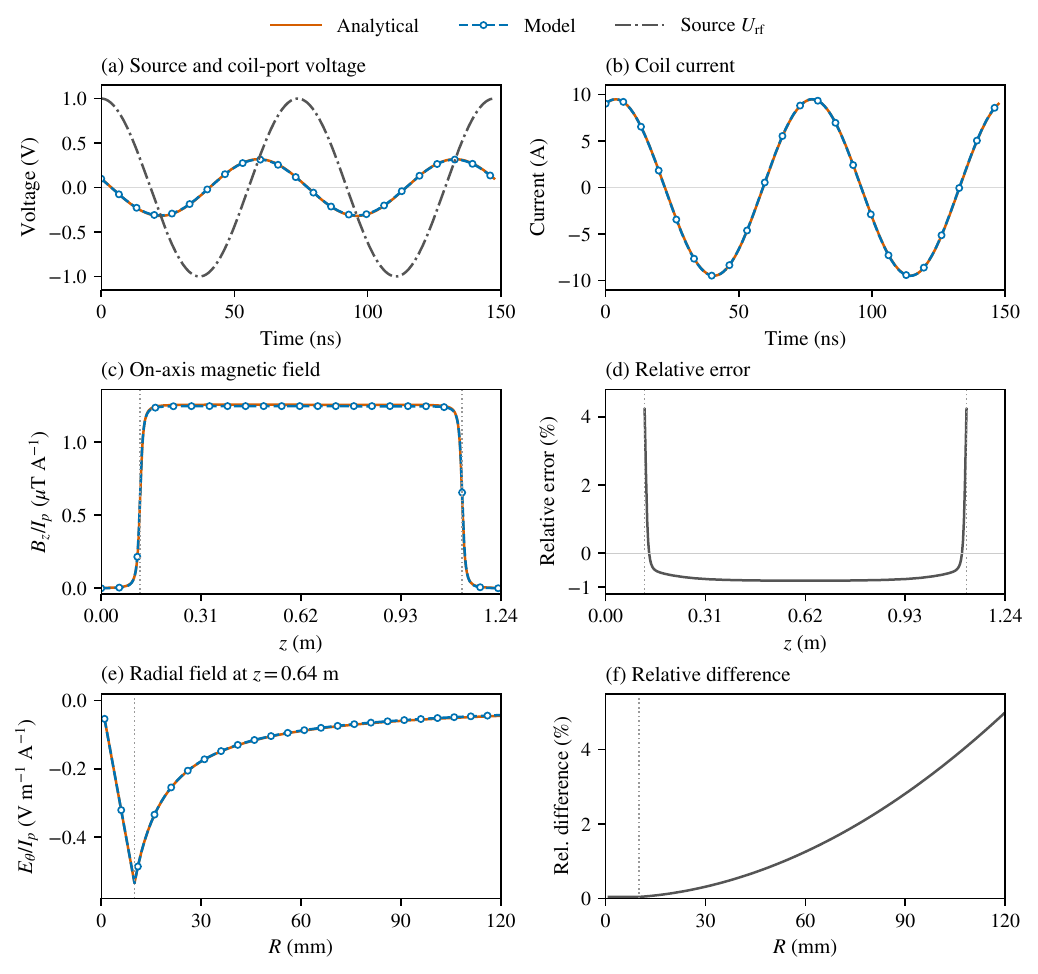}
\caption{Base-case single-solenoid verification against analytical
references. (a)~Source and coil-port voltages and (b)~coil current,
compared with the series-circuit solution. Waveforms use the
$\exp(+\mathrm{i}\omega t)$ convention and peak amplitudes; quantitative
phasors are listed in Table~\ref{tab:S1results}.
(c)~On-axis $B_z/I_p$ at $R=0$ compared with the finite-length
solution~\eqref{eq:S1finiteBz}; (d)~signed relative error,
$100(B_{z,\mathrm{model}}-B_{z,\mathrm{ana}})/B_{z,\mathrm{ana}}$,
reported only over the current-sheet span. Vertical dotted lines in
(c,d) mark the two sheet ends.
(e)~Radial $E_\theta/I_p$ at $z=0.64$~m
compared with the infinite-solenoid analytical solution~\eqref{eq:sol_field};
(f)~absolute relative difference, $100|E_{\theta,\mathrm{model}}-
E_{\theta,\mathrm{ana}}|/|E_{\theta,\mathrm{ana}}|$.
Vertical dotted lines in (e,f) mark the current sheet at $R=a=10$~mm.
The radial comparison concerns the interior
radial structure; the far-field difference is interpreted as an effect of
the finite boundary setting relative to the infinite-domain reference.
No uniform pointwise error bound at the outer boundary is inferred from
this plot.}
\label{fig:S1comparison}
\end{figure*}

Figure~\ref{fig:S1comparison}(a,b) compares the corresponding time-domain
waveforms over approximately two RF periods: panel~(a) shows the source and
coil-port voltages, and panel~(b) shows the coil current. The numerical and
analytical curves are nearly indistinguishable at the plotted scale. The
quantitative comparison uses the phasors listed in
Table~\ref{tab:S1results}. Substitution of the numerical
$Z_{\mathrm{vac}}$ into the algebraic series
circuit reproduces the simulated current with a relative difference of
$5\times10^{-7}$, limited by rounding of the values used in the comparison. The small
analytical--numerical difference is consistent with the $-1.58$\%
field-derived inductance difference, while the internal circuit identity
checks the implementation independently of that field error.

Figure~\ref{fig:S1comparison}(c,d) compares $B_z/I_p$ on $R=0$ with the finite-length
analytical solution~\eqref{eq:S1finiteBz}. The numerical axial value is
obtained by an even-parity extrapolation in $R^2$ from the first three
off-axis grid lines. At the sheet centre, the numerical and analytical
values listed in Table~\ref{tab:S1results} differ by $-0.81$\%. The largest
in-sheet discrepancy is $4.23$\% at the two sheet ends, $z=0.121$ and
$1.120$~m. These localized differences occur where the ideal zero-thickness
sheet terminates and where the finite homogeneous Dirichlet truncation has
its strongest influence; the numerical source is also regularized over two
radial cells. Away from these end regions, the numerical profile closely
follows the analytical solution; the largest differences are therefore
localized at the regularized sheet ends and near the finite-domain boundaries
rather than in the interior field.

Figure~\ref{fig:S1comparison}(e,f) compares the radial distribution of
$E_\theta/I_p$ at $z=0.64$~m with the infinite-solenoid reference in
Eq.~\eqref{eq:sol_field}. In and near the current sheet at $R=a=10$~mm,
the numerical profile follows the linear interior branch and its transition
to the exterior $1/R$ trend. The comparison with the infinite-domain
reference is affected by the finite boundary setting: the numerical problem
imposes $\hat E_\theta(z,b)=0$ through Eq.~\eqref{eq:S1bc}, whereas the
analytical solution retains a nonzero tail at that radius. The increasing
far-field difference is therefore interpreted as a finite-domain truncation
effect. The two boundary-value problems differ, so this comparison does not
require uniform pointwise agreement up to the truncation boundary.
For the present single-port verification, this boundary-setting difference
is acceptable at the reported accuracy of the integral field--circuit
response. The boundary scan in Table~\ref{tab:S1cases} quantifies its
influence on the extracted inductance. Figure~\ref{fig:S1comparison}(e,f) is used to
assess the interior radial structure, without assigning a uniform relative
error bound to the full plotted interval.

For the base vacuum case, the two-dimensional field topology and the
$R=0$ and $z=0.64$~m profiles reproduce the corresponding analytical
structure in the interior region. Differences from the ideal references
include finite-sheet-length, finite-boundary, and discretization effects.
The field-derived inductance differs from the ideal
infinite-solenoid value by $-1.58$\%, while the current-amplitude and
coil-voltage-amplitude differences are $0.16$\% and $1.4$\%, respectively.
Across the boundary, mesh, and sheet-length variations in
Table~\ref{tab:S1cases}, the field-derived inductance error remains below
$3$\% in magnitude and is governed mainly by the finite-boundary truncation,
with an additional finite-length sensitivity. Together, these results verify
the complete single-port inductive path for the tested vacuum, axisymmetric,
single-frequency configuration and establish the diagonal field--circuit
response used by the subsequent tests.

\FloatBarrier
\subsection{Test 2: two coaxial ideal solenoids}
\label{subsec:verif_mutual}

Building on the diagonal response verified in Test~1, the present test adds a
second coaxial port and obtains the off-diagonal impedance from the same field
projection~\eqref{eq:zvac}. It therefore tests the complete mutual-inductive
path, from flux linkage and the projected impedance matrix to the passive
secondary response of the two-port circuit.

\begin{figure*}[!htbp]
\centering
\includegraphics[width=\textwidth]{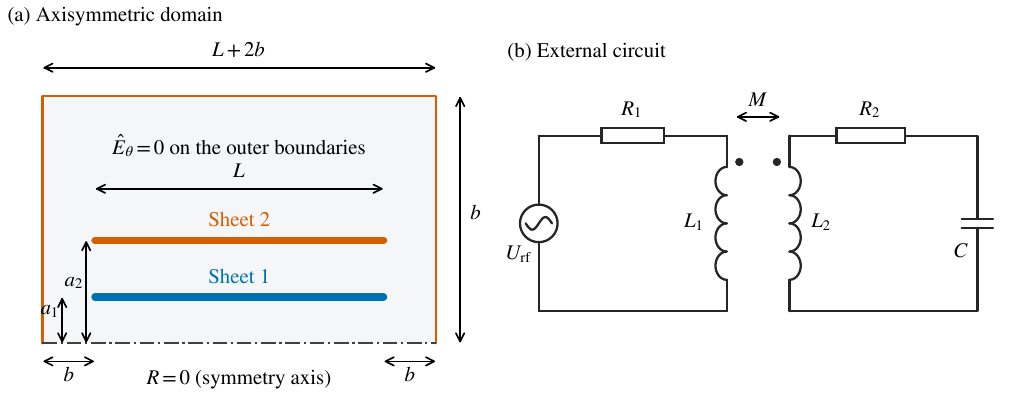}
\caption{Two-solenoid mutual-inductance test (not to scale).
(a)~Axisymmetric vacuum domain containing coaxial azimuthal current sheets
at $R=a_1$ and $R=a_2$. The $R=0$ edge is the symmetry axis, and the other
three edges are finite homogeneous Dirichlet boundaries for
$\hat E_\theta$. (b)~Electrically isolated primary and secondary loops.
The primary contains $U_{\mathrm{rf}}$, $R_1$, and $L_1$; the unpowered
secondary contains $L_2$, $R_2$, and $C$. Their only coupling is the
field-derived mutual impedance
$Z_{\mathrm{vac},12}=\mathrm{i}\omega M$.}
\label{fig:S2model}
\end{figure*}

Figure~\ref{fig:S2model}(a) shows two thin cylindrical current sheets with
radii $a_1$ and $a_2$. Both sheets have length $L$, occupy
$b\le z\le b+L$, and carry the ampere-turn current $\hat I_k$ with
$K_{\theta,k}=\hat I_k/L$. They are internal volume-current sources rather
than field boundaries. The $R=0$ edge is the symmetry axis, where
regularity requires $\hat E_\theta=0$; homogeneous Dirichlet conditions are
imposed at $z=0$, $z=L+2b$, and $R=b$, as in Eq.~\eqref{eq:S1bc}.

The complete model and circuit conditions for this configuration are listed
in Table~\ref{tab:S2conditions}.

\begin{table*}[!tbp]
\centering
\caption{Model and numerical conditions for the two-solenoid test.}
\label{tab:S2conditions}
\begin{tabular}{
>{\raggedright\arraybackslash}p{0.29\textwidth}
>{\raggedright\arraybackslash}p{0.63\textwidth}}
\toprule
item & specification\\
\midrule
Current-sheet geometry
& $a_1=10$~mm, $a_2=20$~mm, $L=1.0$~m, $b=0.12$~m;
both sheets occupy $b\le z\le b+L$\\
Vacuum domain and mesh
& $0\le z\le L+2b$, $0\le R\le b$;
$\Delta z=\Delta R=h=1$~mm;
$N_z\times N_R=1241\times121$\\
Field model and boundaries
& magnetoquasistatic vacuum model without particles or plasma back-emf;
axis regularity and the three finite homogeneous Dirichlet boundaries in
Eq.~\eqref{eq:S1bc}\\
RF sampling
& $f=13.56$~MHz; $N_{\mathrm{RF}}=400$ steps per period; peak-phasor
convention $\exp(+\mathrm{i}\omega t)$\\
Primary loop
& $\hat U_{\mathrm{rf}}=1\angle0^\circ$~V;
$R_1=0.1~\Omega$; zero coil resistance\\
Passive secondary loop
& no independent source; $R_2=0.1~\Omega$; $C=10$~nF
[$1/(\omega C)=1.1737~\Omega$]\\
Coupling and controls
& loops coupled only through the field-derived
$Z_{\mathrm{vac},12}=\mathrm{i}\omega M$;
current under-relaxation factor $\alpha=1$; power control disabled\\
\bottomrule
\end{tabular}
\end{table*}

The two current sheets are connected to the electrically isolated loops in
Fig.~\ref{fig:S2model}(b). In vacuum, the plasma back-emf vanishes, and the
peak-phasor loop equations are
\begin{equation}
\hat U_{\mathrm{rf}}=Z_A\hat I_1+Z_m\hat I_2,\qquad
0=Z_m\hat I_1+Z_B\hat I_2
\label{eq:S2loops}
\end{equation}
\begin{equation}
Z_A=R_1+Z_{\mathrm{vac},11},\qquad
Z_B=Z_{\mathrm{vac},22}+R_2-\frac{\mathrm{i}}{\omega C},\qquad
Z_m=Z_{\mathrm{vac},12}
\label{eq:S2zab}
\end{equation}
with the analytical solution
\begin{equation}
\hat I_1^{\mathrm{ana}}
=\frac{\hat U_{\mathrm{rf}} Z_B}{Z_A Z_B-Z_m^2},\qquad
\hat I_2^{\mathrm{ana}}
=\frac{-Z_m\hat U_{\mathrm{rf}}}{Z_A Z_B-Z_m^2}
\label{eq:S2solution}
\end{equation}
The coil-terminal voltages plotted below are
\begin{equation}
\hat U_1=Z_{\mathrm{vac},11}\hat I_1+
         Z_{\mathrm{vac},12}\hat I_2,\qquad
\hat U_2=Z_{\mathrm{vac},21}\hat I_1+
         Z_{\mathrm{vac},22}\hat I_2
\label{eq:S2ports}
\end{equation}

For two infinitely long coaxial thin current sheets, the analytical
inductances are
\begin{equation}
\begin{aligned}
L_1^{\mathrm{ana}}=\mu_0\frac{\pi a_1^2}{L},\qquad
L_2^{\mathrm{ana}}=\mu_0\frac{\pi a_2^2}{L},\\
M^{\mathrm{ana}}=\mu_0\frac{\pi a_1^2}{L},\qquad
k^{\mathrm{ana}}=\frac{M^{\mathrm{ana}}}
{\sqrt{L_1^{\mathrm{ana}}L_2^{\mathrm{ana}}}}=\frac{a_1}{a_2}
\end{aligned}
\label{eq:S2ind}
\end{equation}
because the axial field of an infinite sheet fills its whole bore, so that
the flux linked by the inner sheet is limited by its own cross-section
$\pi a_1^2$. The numerical mutual inductance is extracted directly from the
off-diagonal field projection,
\begin{equation}
M_{\mathrm{num}}
=\frac{\operatorname{Im}Z_{\mathrm{vac},12}}{\omega}
\label{eq:S2Mnum}
\end{equation}

\begin{table}[!htbp]
\centering
\small
\setlength{\tabcolsep}{3pt}
\caption{Analytical and model results for the two-solenoid test. Current and
voltage phasors are peak values in polar form under the
$\exp(+\mathrm{i}\omega t)$ convention.}
\label{tab:S2results}
\begin{tabular}{lcc}
\toprule
quantity & analytical solution & model result\\
\midrule
$L_1$ [nH] & 0.394784 & 0.388555\\
$L_2$ [nH] & 1.579137 & 1.513725\\
$M$ [nH] & 0.394784 & 0.377497\\
$k$ & 0.500000 & 0.492225\\
$\hat I_1$ [A]
& $9.43822\angle(-19.1257^\circ)$
& $9.45728\angle(-18.8058^\circ)$\\
$\hat I_2$ [A]
& $0.304090\angle(-24.6224^\circ)$
& $0.289822\angle(-24.2734^\circ)$\\
$\hat U_1$ [V]
& $0.327643\angle70.7030^\circ$
& $0.322363\angle71.0364^\circ$\\
$\hat U_2$ [V]
& $0.358207\angle70.2474^\circ$
& $0.341399\angle70.5965^\circ$\\
\bottomrule
\end{tabular}
\end{table}

\begin{figure*}[!htbp]
\centering
\includegraphics[width=\textwidth]{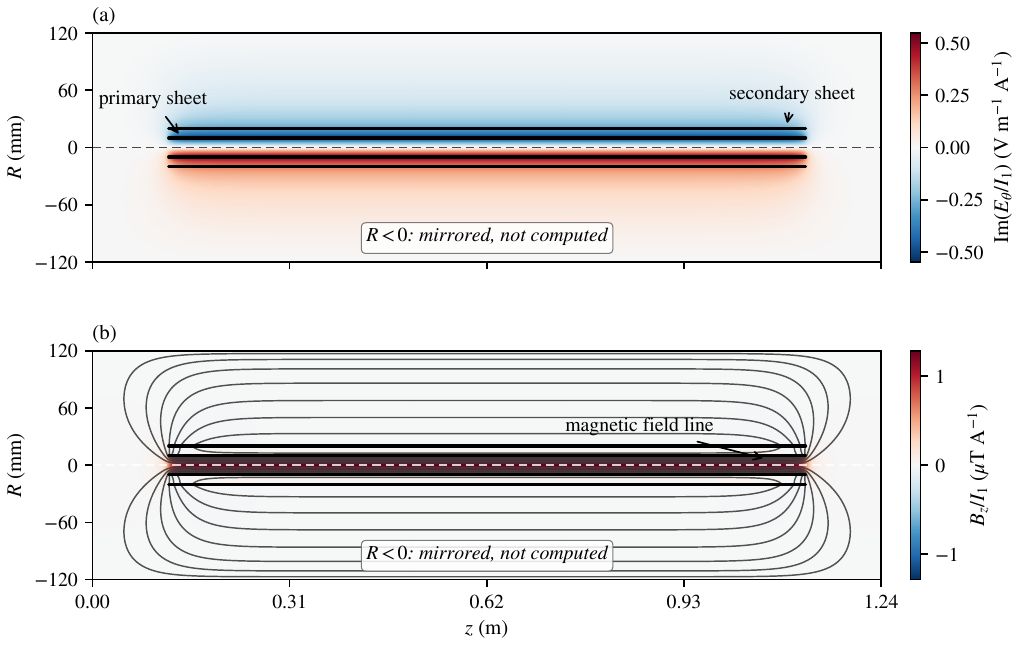}
\caption{Two-dimensional field for the two-solenoid test.
(a)~Primary-current-referenced reactive azimuthal field
$\operatorname{Im}(\hat E_\theta/\hat I_1)$, with the primary and secondary
current sheets marked. (b)~Associated axial magnetic field $B_z/I_1$ and
magnetic-field lines. Only $R\ge0$ is computed; $R<0$ is the axisymmetric
mirror used for visualization.}
\label{fig:S2fields}
\end{figure*}

Figure~\ref{fig:S2fields} shows the two-dimensional field for this
configuration. The upper half-plane, $0\le R\le120$~mm, is the computed
domain. The negative-$R$ half is included only as an axisymmetric
visualization: $\operatorname{Im}(\hat E_\theta/\hat I_1)$ is extended with
odd parity and $B_z/I_1$ with even parity. The field is not fitted to obtain
$M$; the mutual term follows from the source--test projection
\eqref{eq:S2Mnum}. The magnetic-field lines pass through the bores of both
coaxial sheets and therefore provide the flux linkage represented by the
off-diagonal impedance.

The analytical--model comparison is collected in
Table~\ref{tab:S2results}. For the $b=0.12$~m, $h=1$~mm discretization, the
field-derived mutual inductance is $4.38$\% lower than the
infinite-solenoid reference. The difference is consistent with the combined
finite-sheet-length, finite-boundary, and spatial-discretization effects
identified by the Test~1 sensitivity study.

\begin{figure*}[!htbp]
\centering
\includegraphics[width=\textwidth]{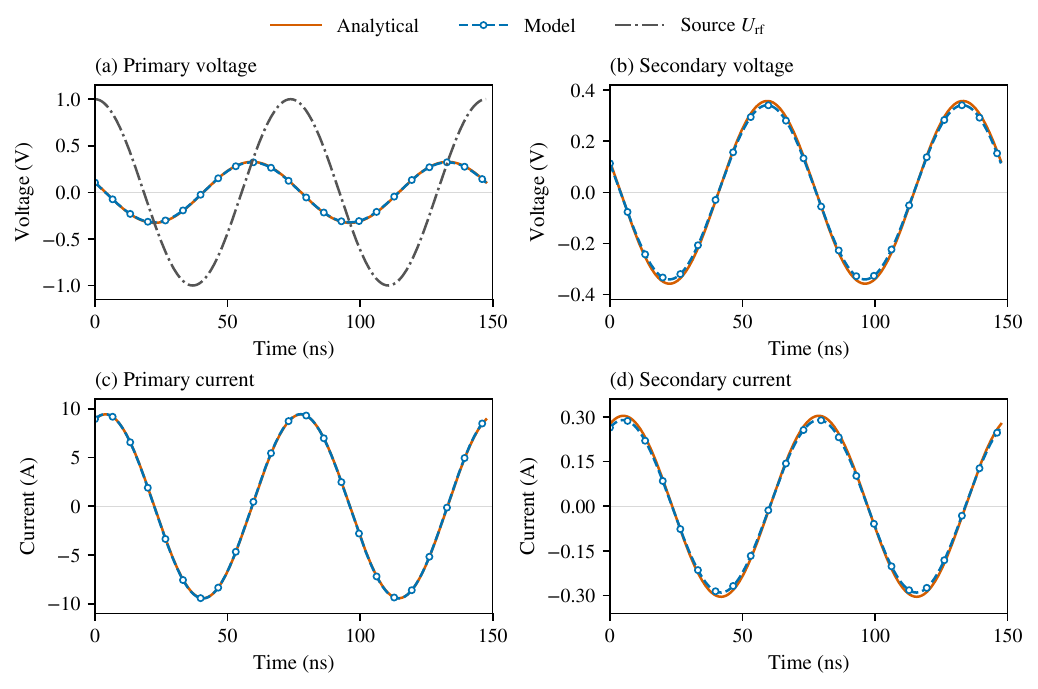}
\caption{Analytical and model waveforms for the two-solenoid configuration.
(a)~Source and primary coil-terminal voltage; (b)~secondary coil-terminal
voltage; (c)~primary current; (d)~secondary current. The curves are
reconstructed from peak phasors using the $\exp(+\mathrm{i}\omega t)$
convention; the quantitative phasors are listed in
Table~\ref{tab:S2results}.}
\label{fig:S2waveforms}
\end{figure*}

Figure~\ref{fig:S2waveforms} compares the analytical and model port
waveforms over approximately two RF periods. Panels~(a) and~(b) show the
primary and secondary coil-terminal voltages, respectively, whereas
panels~(c) and~(d) show the corresponding currents. Using the full complex
phasors in Table~\ref{tab:S2results}, the relative differences are $0.594$\%
for $\hat I_1$,
$1.712$\% for $\hat U_1$, and $4.730$\% for both $\hat I_2$ and
$\hat U_2$. The secondary loop contains no source. Its nonzero voltage and
current therefore arise from $Z_{\mathrm{vac},21}=\mathrm{i}\omega M$ in
Eqs.~\eqref{eq:S2loops} and~\eqref{eq:S2ports}. Agreement of this passive
secondary response with the analytical two-loop solution checks that the
field-derived mutual term is not only present in the impedance matrix but
is also used with the correct magnitude and phase by the circuit solver.

In the tested vacuum configuration, the off-diagonal field projection gives
a mutual inductance within $4.38$\% of the infinite-solenoid reference, and
the resulting passive-secondary voltage and current remain within $4.73$\%
of the analytical two-loop response. These results verify the magnitude and
phase transfer of the mutual-inductive coupling through the field solver,
impedance matrix, and circuit solution.

\FloatBarrier
\subsection{Test 3: joint inductive--capacitive coupling in a five-node
manufactured benchmark}
\label{subsec:verif_joint_lc}

This test examines the simultaneous closure of the inductive and capacitive
channels in the distributed circuit. Five nested cylindrical sheets activate
the full mutual-inductance matrix, spatial capacitive coupling, and all eleven
complex network unknowns. The calculation uses the field, surface-charge,
and circuit routines of the PIC/MCC code, including the
$J_\theta\mathbin{\to}\hat E_\theta$ inductive response of
Sec.~\ref{subsec:inductive}, with manufactured boundary
conditions. Particles are absent, so comparison with an analytical network isolates the linear
field--circuit coupling from plasma kinetics. The plasma reaction voltage
and particle energy deposition vanish in this test. The physical and numerical
settings are listed in Table~\ref{tab:t7parameters}.

\begin{figure*}[!htbp]
\centering
\includegraphics[width=\textwidth]{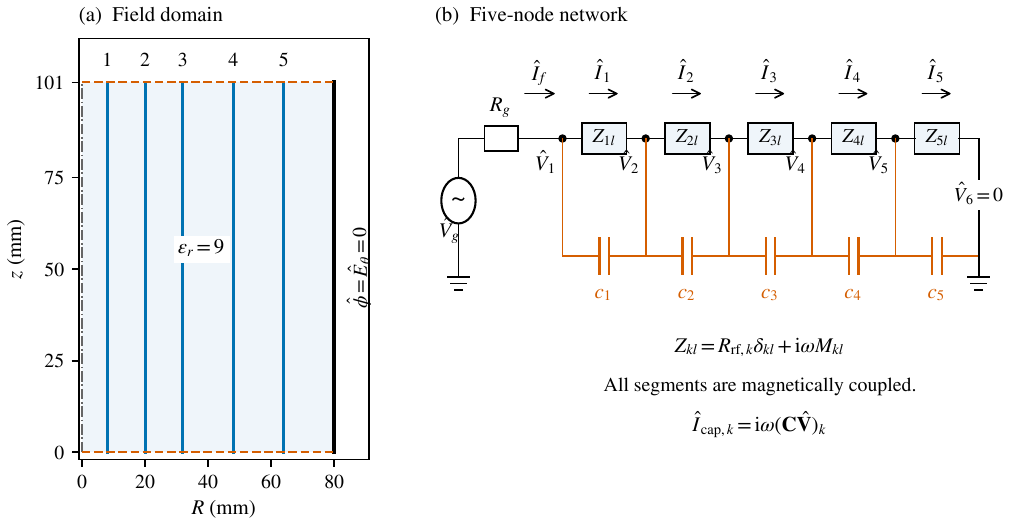}
\caption{Five-node manufactured benchmark. (a)~Meridional computational
domain for $h=1$~mm. The five internal blue lines mark the cylindrical
sheets, numbered in electrical order; the dashed axial boundaries carry
manufactured Dirichlet values. The left boundary is the symmetry axis, and
the right boundary is the grounded outer radius. Here $\ell=100$~mm is the
source-normalization length and $L_z=101$~mm is the grid length.
(b)~Circuit topology, drawn schematically. Each voltage node precedes its
series segment, and the endpoint after segment~5 is grounded; $\hat V_5$
remains an unknown. The orange capacitors represent the analytical gap
capacitances $c_k$; the numerical calculation obtains their combined response
from the spatial field. Each series box denotes one row of the coupled
impedance matrix, so its voltage drop depends on all five segment currents.}
\label{fig:t7geometry}
\end{figure*}

Figure~\ref{fig:t7geometry} shows the axisymmetric domain and its electrical
topology. Sheet~$k$ has radius $a_k$, uniform azimuthal sheet-current
density $\hat K_{\theta,k}=\hat I_k/\ell$, and electrostatic potential
$\hat V_k$. The dielectric is homogeneous and lossless. At the axis,
$\partial_R\hat\phi=0$ and $\hat E_\theta=0$; the outer boundary at $R=b$
satisfies $\hat\phi=\hat E_\theta=0$. Both axial boundaries take the
corresponding analytical radial solutions given in
Appendix~\ref{app:joint_lc_details}. For the electrostatic problem,
these boundary values are reconstructed from the circuit voltages applied
during the current RF period. Thus, the boundary treatment does not prescribe
the converged circuit state. It removes physical end effects from the
continuous reference problem, while retaining the two-dimensional field
solves. The effective source length is $\ell=N_{\mathrm{src}}h$; the
boundary-to-boundary grid length is $L_z=\ell+h$.

\begin{table*}[!tbp]
\centering
\caption{Physical and numerical parameters of the joint benchmark. Voltage
and current amplitudes are peak phasors under
$\operatorname{Re}[\hat x\exp(+\mathrm{i}\omega t)]$.}
\label{tab:t7parameters}
\small
\begin{tabular}{@{}p{0.38\textwidth}p{0.57\textwidth}@{}}
\toprule
Parameter & Value or setting \\
\midrule
Sheet radii $a_1,\ldots,a_5$ & $8,20,32,48,64$~mm \\
Outer radius; effective length & $b=80$~mm; $\ell=100$~mm \\
Relative material constants & $\varepsilon_r=9$, $\mu_r=1$; no particles \\
Constants used in code and reference &
$\varepsilon_0=8.8542\times10^{-12}$~F\,m$^{-1}$;
$\mu_0=1.2566\times10^{-6}$~H\,m$^{-1}$ \\
Source & $f=13.56$~MHz; $\hat V_g=100\angle0^\circ$~V; $R_g=50~\Omega$ \\
Segment resistance & $R_{\mathrm{rf},k}=20~\Omega$ for $k=1,\ldots,5$ \\
Other circuit elements & No matching elements or external stray capacitance \\
Principal grid & $h=\Delta R=\Delta z=1$~mm; $(N_z,N_R)=(102,81)$ \\
Spatial refinement & $h=2,1,0.5,0.25$~mm, with fixed $\ell$ \\
Period update & 400 samples per RF period; 30 completed periods; $\alpha=1$ \\
Field solves and parallelism & MQS inductive operator; Laplace electrostatics;
GMRES relative tolerance $10^{-12}$; 8 MPI ranks \\
Control settings & Fixed source voltage; power controller disabled \\
\bottomrule
\end{tabular}
\end{table*}

The continuous reference follows from the shielded cylindrical field
solutions. With $a_6=b$ and $c_0=0$, the gap capacitances and nonzero entries
of the Maxwell capacitance matrix are
\begin{equation}
\begin{aligned}
c_k&=\frac{2\pi\varepsilon_0\varepsilon_r\ell}
              {\ln(a_{k+1}/a_k)},\qquad k=1,\ldots,5,\\
C_{kk}&=c_{k-1}+c_k,\qquad
C_{k,k+1}=C_{k+1,k}=-c_k\quad(k<5)
\end{aligned}
\label{eq:t7C}
\end{equation}
All other entries vanish. For identically oriented current sheets, the
self- and mutual-inductance matrix is
\begin{equation}
M_{kl}=\frac{\mu_0\pi}{\ell}\min(a_k,a_l)^2
       \left[1-\frac{\max(a_k,a_l)^2}{b^2}\right]
\label{eq:t7M}
\end{equation}
The finite-$b$ factor enforces the outer inductive boundary condition and
must be retained in the reference. This manufactured magnetoquasistatic (MQS)--electroquasistatic (EQS)
problem isolates the field--circuit algorithm under prescribed axial
boundaries; physical coil end effects and wave propagation are excluded.

The unknown vector is
$\mathbf x=(\hat I_f,\hat I_1,\ldots,\hat I_5,
\hat V_1,\ldots,\hat V_5)^{\mathrm T}$. In steady state it satisfies
\begin{equation}
\begin{aligned}
R_g\hat I_f+\hat V_1&=\hat V_g,\\
\hat I_{\mathrm{in},k}-\hat I_k
 &=\mathrm{i}\omega(\mathbf C\hat{\mathbf V})_k,\\
\hat V_k-\hat V_{k+1}
 &=R_{\mathrm{rf},k}\hat I_k+
   \mathrm{i}\omega\sum_{l=1}^{5}M_{kl}\hat I_l,
\end{aligned}
\label{eq:t7network}
\end{equation}
where $\hat I_{\mathrm{in},1}=\hat I_f$,
$\hat I_{\mathrm{in},k}=\hat I_{k-1}$ for $k>1$, and $\hat V_6=0$.
Positive capacitive current leaves the voltage node and enters the spatial
domain. The continuous solution $\mathbf x_*$ is obtained by solving these
eleven equations independently of the field code.

The numerical calculation determines $\mathbf M_h$ by projecting the five
unit-current field solutions onto the current-sheet distributions. Five
additional Laplace basis runs, each applying 1~V to one sheet and zero to
the others, provide $\mathbf C_h$ through the surface-charge integral.
The field-derived $\mathbf M_h$ supplies the inductive impedance of the
coupled calculation. Together with the separately measured $\mathbf C_h$,
it also defines an independent semidiscrete network reference
$\mathbf x_{h,*}$ through Eq.~\eqref{eq:t7network}. The basis-run
capacitance matrix is not supplied to the coupled solver. During that run, the
capacitive current is obtained from the time-sampled charge and its
fundamental Fourier component, $\hat{\mathbf I}_{\mathrm{cap}}
=\mathrm{i}\omega\hat{\mathbf Q}$. For this benchmark, the vacuum field
solver was extended to solve Laplace's equation with circuit-driven
conductor potentials. The results reported here apply to this
implementation. Each field solve uses the GMRES tolerance in
Table~\ref{tab:t7parameters} and terminates the calculation if convergence
is not achieved.

\begin{figure*}[!htbp]
\centering
\includegraphics[width=\textwidth]{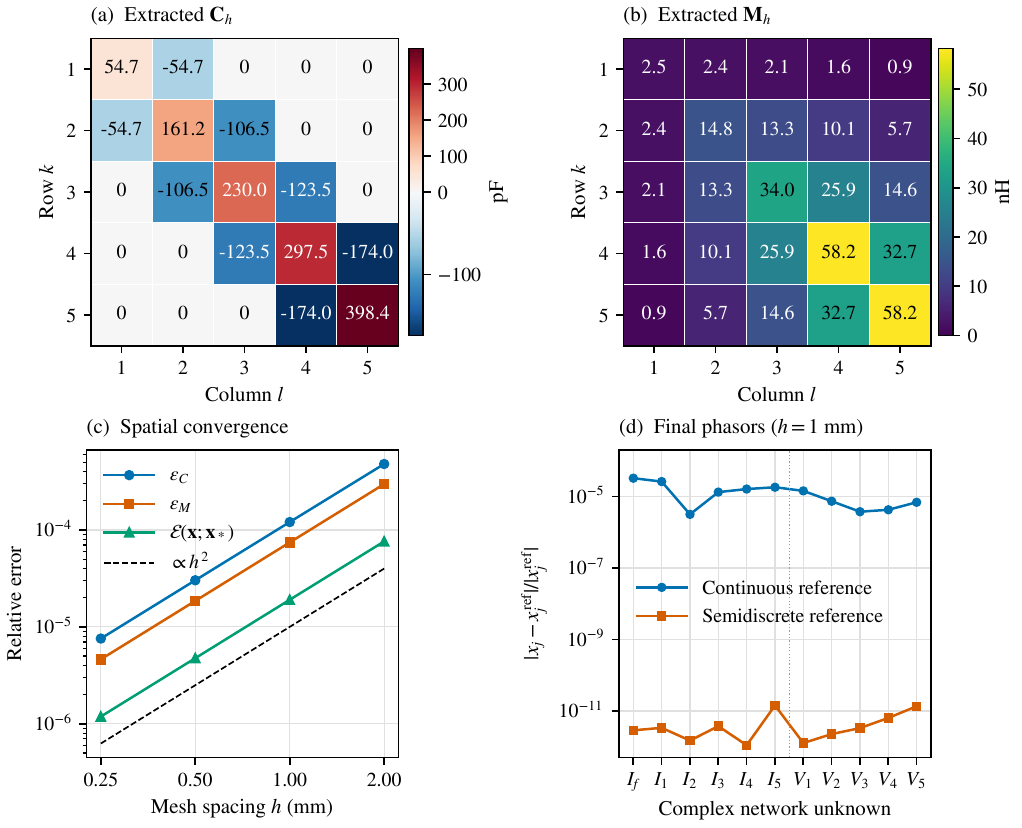}
\caption{Field extraction and steady-state accuracy. (a,b)~Numerically
extracted $\mathbf C_h$ in pF and $\mathbf M_h$ in nH at $h=1$~mm; matrix
indices follow the node order in Fig.~\ref{fig:t7geometry}. Printed entries
are rounded for display. (c)~Relative Frobenius errors of the two matrices
and scaled state error against the continuous reference on four grids.
The dashed line indicates second-order scaling, with arbitrary vertical
normalization. (d)~Relative complex-phasor errors for all eleven unknowns
after 30 completed RF periods. Circles use the continuous reference;
squares use the independent network constructed from $\mathbf C_h$ and
$\mathbf M_h$. Lines between discrete points are visual guides.}
\label{fig:t7accuracy}
\end{figure*}

The two references distinguish field-discretization error from errors in
network assembly and feedback. Matrix errors are measured with the relative
Frobenius norm. For the mixed current--voltage state, we use
\begin{equation}
\mathcal E(\mathbf x;\mathbf x^{\mathrm{ref}})=
\frac{\|\mathbf S^{-1}(\mathbf x-\mathbf x^{\mathrm{ref}})\|_2}
     {\|\mathbf S^{-1}\mathbf x^{\mathrm{ref}}\|_2},\qquad
\mathbf S=\operatorname{diag}
\bigl(|\hat I_f^{\mathrm{ref}}|\mathbf 1_6,
       |\hat V_g|\mathbf 1_5\bigr)
\label{eq:t7error}
\end{equation}
At $h=1$~mm, the relative errors of $\mathbf C_h$ and $\mathbf M_h$ are
$1.21\times10^{-4}$ and $7.44\times10^{-5}$, respectively
(Fig.~\ref{fig:t7accuracy}). After 30 RF periods, the final network state
has errors $1.90\times10^{-5}$ against $\mathbf x_*$ and
$6.26\times10^{-12}$ against $\mathbf x_{h,*}$. The largest relative
error of an individual complex phasor against the continuous reference is
$3.26\times10^{-5}$. These comparisons include both phase and amplitude.
Refining $h$ from 2 to 0.25~mm reduces the continuous-reference state error
from $7.63\times10^{-5}$ to $1.19\times10^{-6}$, with adjacent-grid orders
of approximately 2.00. The matrix errors show the same second-order trend.
The semidiscrete state errors remain below $9.1\times10^{-11}$ on all four
grids.

\begin{figure*}[!htbp]
\centering
\includegraphics[width=\textwidth]{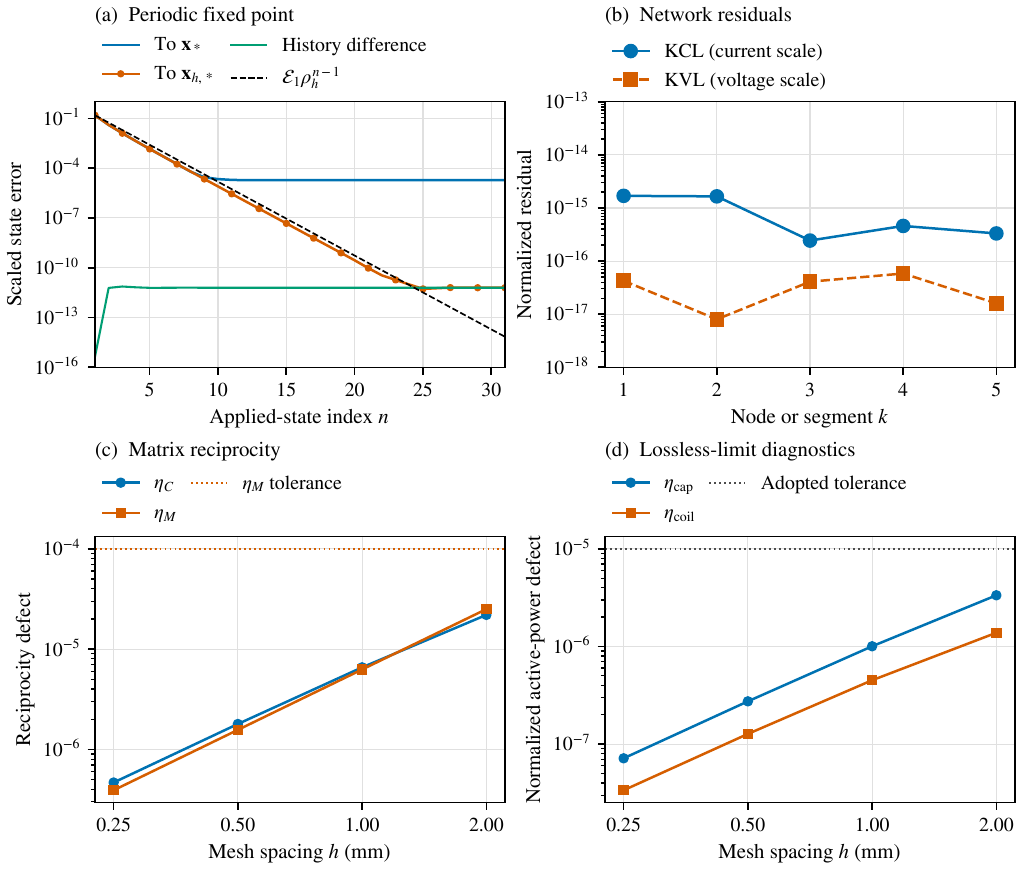}
\caption{Periodic feedback and finite-grid defects. (a)~Scaled distances
to the continuous and semidiscrete fixed points, and the difference between
the computed history and the semidiscrete recurrence. State~1 is the
zero-capacitive-feedback initialization; state~31 is prepared after cycle~30
and is not an additional simulated period. The dashed spectral-radius
curve indicates an asymptotic rate, not a bound on every iterate.
(b)~Normalized final KCL and KVL residuals computed from the assembled
network matrix, right-hand side, and solution. (c)~Reciprocity defects
$\eta_C$ and $\eta_M$, with the adopted $10^{-4}$ tolerance for $\eta_M$.
(d)~Normalized active-power defects $\eta_{\mathrm{cap}}$ and
$\eta_{\mathrm{coil}}$, both with an adopted tolerance of $10^{-5}$.
Dotted horizontal lines indicate these tolerances; symbols are
computed results, and connecting lines are visual guides.}
\label{fig:t7convergence}
\end{figure*}

The cycle-history comparison tests the one-period delay of the implemented
feedback. This history describes the convergence of the partitioned
algorithm, rather than a resolved physical switch-on transient. Let
$\mathbf A_{0,h}$ contain the source and inductive network
terms, with capacitive currents on the right-hand side; let $\mathbf T$
insert these currents into the five KCL rows and $\mathbf S_V$ select the
five voltages. With unit relaxation, the independent recurrence is
\begin{equation}
\begin{aligned}
\mathbf x_h^{n+1}&=\mathbf c_h+\mathbf G_h\mathbf x_h^n,\qquad
\mathbf x_h^1=\mathbf c_h,\\
\mathbf c_h&=\mathbf A_{0,h}^{-1}\mathbf b_0,\qquad
\mathbf G_h=\mathbf A_{0,h}^{-1}\mathbf T
             (\mathrm{i}\omega\mathbf C_h)\mathbf S_V,
\end{aligned}
\label{eq:t7recurrence}
\end{equation}
where $\mathbf b_0=(\hat V_g,0,\ldots,0)^{\mathrm T}$.
The spectral radius is $\rho(\mathbf G_h)=0.359069$, compared with
0.359062 for the continuous matrices. Figure~\ref{fig:t7convergence}(a)
compares the computed and reconstructed histories using the actual
applied-state indices. Their maximum scaled difference is
$7.40\times10^{-12}$, using the fixed scales and denominator associated
with $\mathbf x_{h,*}$ in Eq.~\eqref{eq:t7error}. For indices $n=21$--31,
the maximum distance to this fixed point is $9.47\times10^{-11}$.
The final assembled KCL and KVL residuals are below $1.7\times10^{-15}$
and $6\times10^{-17}$ when normalized by $|\hat I_f|$ and $|\hat V_g|$,
respectively. These algebraic residuals complement the independent
reference comparisons; they do not measure field accuracy.

The power comparison must retain the small nonsymmetric parts of the
extracted matrices. Define the quadratic forms
$W_M=\hat{\mathbf I}^{\mathrm H}\mathbf M_h\hat{\mathbf I}$ and
$W_C=\hat{\mathbf V}^{\mathrm H}\mathbf C_h\hat{\mathbf V}$, where
$\mathrm H$ denotes conjugate transpose. At the coupled fixed point, the network implies
\begin{equation}
S_g=\frac12\hat V_g\hat I_f^*
=P_{R_g}+P_{\mathrm{Cu}}+
  \frac{\mathrm{i}\omega}{2}\left(W_M-W_C^*\right),
\label{eq:t7power}
\end{equation}
with $P_{R_g}=R_g|\hat I_f|^2/2$ and
$P_{\mathrm{Cu}}=\sum_k R_{\mathrm{rf},k}|\hat I_k|^2/2$.
This full complex identity closes to a relative residual of
$2.93\times10^{-12}$ at $h=1$~mm. The inductive and capacitive reactive
contributions are $Q_L=(\omega/2)\operatorname{Re}W_M=7.28812$~var and
$Q_C=-(\omega/2)\operatorname{Re}W_C=-6.01266$~var. Both channels therefore
contribute to the joint solution.

The reciprocity error is
$\eta_M=\|\mathbf M_h-\mathbf M_h^{\mathrm T}\|_F/\|\mathbf M_h\|_F$
and the two normalized active-power errors are
$\eta_{\mathrm{cap}}=|P_{\mathrm{cap}}|/(|Q_L|+|Q_C|)$ and
$\eta_{\mathrm{coil}}=|P_{\mathrm{port}}-P_{\mathrm{Cu}}|
/|P_{\mathrm{port}}|$, using the powers of the final completed
period. The capacitance reciprocity defect
$\eta_C$ is defined analogously to $\eta_M$. On the principal grid,
$\eta_M=6.27\times10^{-6}$, $\eta_{\mathrm{cap}}=1.004\times10^{-6}$, and
$\eta_{\mathrm{coil}}=4.51\times10^{-7}$; their four-grid maxima are listed
in Table~\ref{tab:t7defects}.

At $h=1$~mm, $P_{\mathrm{cap}}=1.34\times10^{-5}$~W, while the particle
inductive deposition is zero. The nonsymmetric matrix contribution to
the source active power is
$-(\omega/2)\operatorname{Im}(W_M+W_C)=9.57\times10^{-6}$~W.
Thus, closure of Eq.~\eqref{eq:t7power} does not imply exact discrete
losslessness. Figure~\ref{fig:t7convergence}(c,d) shows that these defects
decrease with refinement. In particular, $P_{\mathrm{cap}}$ falls to
$9.51\times10^{-7}$~W at $h=0.25$~mm, with successive observed orders
1.74, 1.87, and 1.94. The trend is consistent with a spatial-discretization
effect; it is not evidence of dielectric or particle heating.

The adopted tolerances of $10^{-4}$, $10^{-5}$, and $10^{-5}$ for these
three diagnostics were selected after examining the computed errors, and
the errors on all four grids fall below them; the original thresholds and
their revision are recorded in Appendix~\ref{app:joint_lc_details}.

\begin{table}[!htbp]
\centering
\small
\setlength{\tabcolsep}{3pt}
\caption{Finite-grid errors and adopted tolerances. The maximum
is taken over $h=2,1,0.5,0.25$~mm. All errors and tolerances are
dimensionless. The tolerances were selected after examining the computed
errors.}
\label{tab:t7defects}
\begin{tabular}{@{}lccc@{}}
\toprule
Diagnostic & $h=1$~mm & Four-grid maximum & Tolerance \\
\midrule
$\eta_M$ & $6.27\times10^{-6}$ & $2.50\times10^{-5}$ & $<10^{-4}$ \\
$\eta_{\mathrm{cap}}$ & $1.004\times10^{-6}$ & $3.34\times10^{-6}$ & $<10^{-5}$ \\
$\eta_{\mathrm{coil}}$ & $4.51\times10^{-7}$ & $1.38\times10^{-6}$ & $<10^{-5}$ \\
\bottomrule
\end{tabular}
\end{table}

Channel perturbations supplement the joint-reference comparison.
Removing the inductive matrix changes the scaled state by 22.4\% relative
to the full continuous reference and agrees with the corresponding
semidiscrete solution to $4.99\times10^{-12}$. Reducing
$\varepsilon_r$ from 9 to 1 scales the extracted capacitance by $1/9$
with relative discrepancy $2.58\times10^{-15}$, while the extracted
inductance is unchanged in the numerical comparison. The coupled state for
this lower-permittivity case agrees with its semidiscrete reference to
$6.77\times10^{-13}$. These perturbations check the distinct inductive
and dielectric dependences of the coupled response. Field profiles,
complete phasors, and numerical reproducibility checks are given in
Appendix~\ref{app:joint_lc_details}.

The evidence supports the coupled field extraction, complex network
assembly, and period-delayed feedback in this particle-free linear
MQS--EQS setting, with the finite-grid errors reported in
Table~\ref{tab:t7defects}. The test does not assess
three-dimensional winding or end effects, wave propagation, nonlinear
plasma loading, nonzero numerical susceptibility $\chi$ or plasma reaction
voltages, particle heating
and noise, higher harmonics, or power-controller stability.
\FloatBarrier

\subsection{Representative application: two-dimensional GEC reference cell}
\label{subsec:application}

\begin{figure*}[!htbp]
\centering
\includegraphics[width=\textwidth]{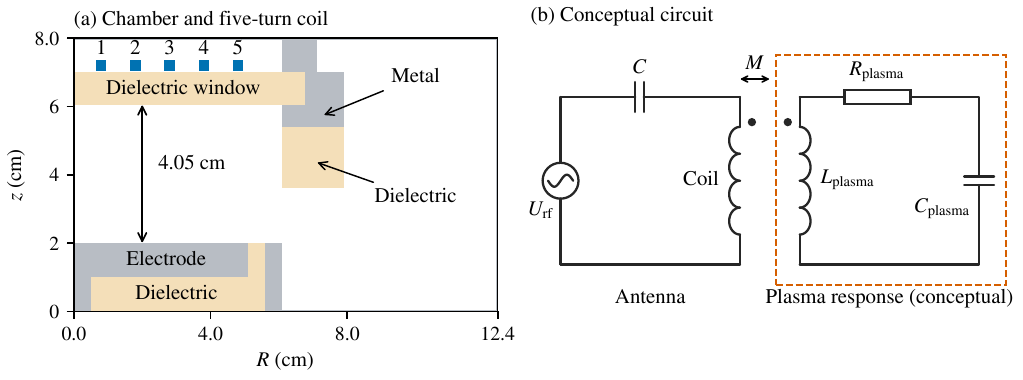}
\caption{GEC application model. (a)~Axisymmetric chamber and five-turn coil
geometry resolved by the PIC/MCC calculation. (b)~Conceptual equivalent
circuit illustrating inductive and capacitive coil--plasma coupling. The
plasma branch in panel (b) is an interpretation only; the plasma response
used by the circuit is obtained self-consistently from the spatial PIC/MCC
solution in panel (a).}
\label{fig:GECmodel}
\end{figure*}

The GEC RF reference cell was introduced as a defined geometry for
cross-laboratory discharge experiments~\cite{hargis1994gaseous}. An
inductively coupled source was subsequently developed for this platform
\cite{miller1995inductively}, and two-dimensional modeling and external
electrical measurements established complementary reference data for the ICP
configuration~\cite{lymberopoulos1995two,singh2008electrical}. The complete
coupling algorithm was applied here to a five-turn coil surrounding an
axisymmetric GEC reference cell. Figure~\ref{fig:GECmodel}(a) shows the
resolved chamber geometry, whereas Fig.~\ref{fig:GECmodel}(b) provides a
conceptual circuit interpretation of the coupled energy-transfer paths. In
the calculation, the GEC chamber and its plasma are resolved directly: the
charged-particle kinetics, space charge, electrostatic potential, and induced
azimuthal field are advanced self-consistently by PIC/MCC, and the resulting
plasma current and turn-resolved surface charge are returned to the external
circuit. The updated circuit solution then supplies the coil currents and
turn potentials for the next RF cycle. Thus, neither the plasma density nor a
lumped plasma impedance is prescribed; in particular, the illustrative
$R_{\mathrm{plasma}}$--$L_{\mathrm{plasma}}$--$C_{\mathrm{plasma}}$ branch
in Fig.~\ref{fig:GECmodel}(b) is not substituted for the PIC/MCC plasma.

The principal conditions are listed in Table~\ref{tab:GECconditions}. Pure
argon was simulated at 13.56~MHz and 10~mTorr (1.333~Pa). The lower electrode
and chamber wall were grounded and no RF bias was applied. The target loaded
port powers were 60 and 100~W. Both cases were advanced for 2000 RF cycles
to obtain the final statistically stationary plasma--circuit operating
points. The results below include the transient circuit histories and final
operating-point comparisons. The circuit and
power quantities were averaged over cycles 1901--2000, after the plasma
response and circuit operating point had settled in the sense defined in
Sec.~\ref{subsec:circuit}. This window samples the final statistical steady
state, rather than defining convergence solely by attainment of the target
port power. A separate vacuum calculation was used only to identify
the coil response in the absence of plasma.

\begin{table*}[!tbp]
\centering
\caption{Principal conditions for the two-dimensional GEC application.}
\label{tab:GECconditions}
\small
\begin{tabular}{@{}>{\raggedright\arraybackslash}p{4.2cm}p{9.0cm}@{}}
\toprule
quantity & value\\
\midrule
chamber domain & $0\le R\le12.4$~cm, $0\le z\le8.0$~cm\\
electrode gap & 4.05~cm\\
gas and pressure & Ar, 10~mTorr (1.333~Pa)\\
RF frequency & 13.56~MHz\\
coil and matching element & five stacked axisymmetric turns, 200~pF series
capacitor; no Faraday shield\\
electrical boundaries & grounded lower electrode and wall; no RF bias\\
mesh and time step & $N_R\times N_z=497\times321$ grid points,
$\Delta R=\Delta z=0.25$~mm;\\
& $\Delta t=1/(400f)\approx1.8436578\times10^{-10}$~s (exactly 400 steps per RF cycle)\\
field operators & inductive operator $\mathcal L$ of
Eq.~\eqref{eq:helmholtz_phasor} with the displacement-current term retained;
Poisson solve with implicit susceptibility\\
initial charged-particle density & $n_{e}=n_{\mathrm{Ar}^{+}}=1.0\times10^{16}$~m$^{-3}$\\
PIC/MCC settings & implicit electrostatic advance; 40 computational particles
per cell at initialization; neutral MCC, secondary-electron emission, and
particle split--merge enabled; Coulomb collisions and recombination disabled\\
loaded operating points & $P_{\mathrm{target}}=60$ and 100~W\\
reported steady-state window & RF cycles 1901--2000\\
\bottomrule
\end{tabular}
\end{table*}

The underlying direct-implicit axisymmetric PIC/MCC model has been reported
previously~\cite{chen2024gec,chen2025eh}.
Neutral collisions use the standard
null-collision construction~\cite{vahedi1995mcc}, and the split--merge module
uses the stochastic particle-control scheme developed for this code
family~\cite{chen2025swpc}. These inherited particle modules are held fixed
in the present application; the new element being exercised is the
turn-resolved coil--field--circuit coupling.

\begin{figure}[!htbp]
\centering
\includegraphics[width=\figw{0.90}]{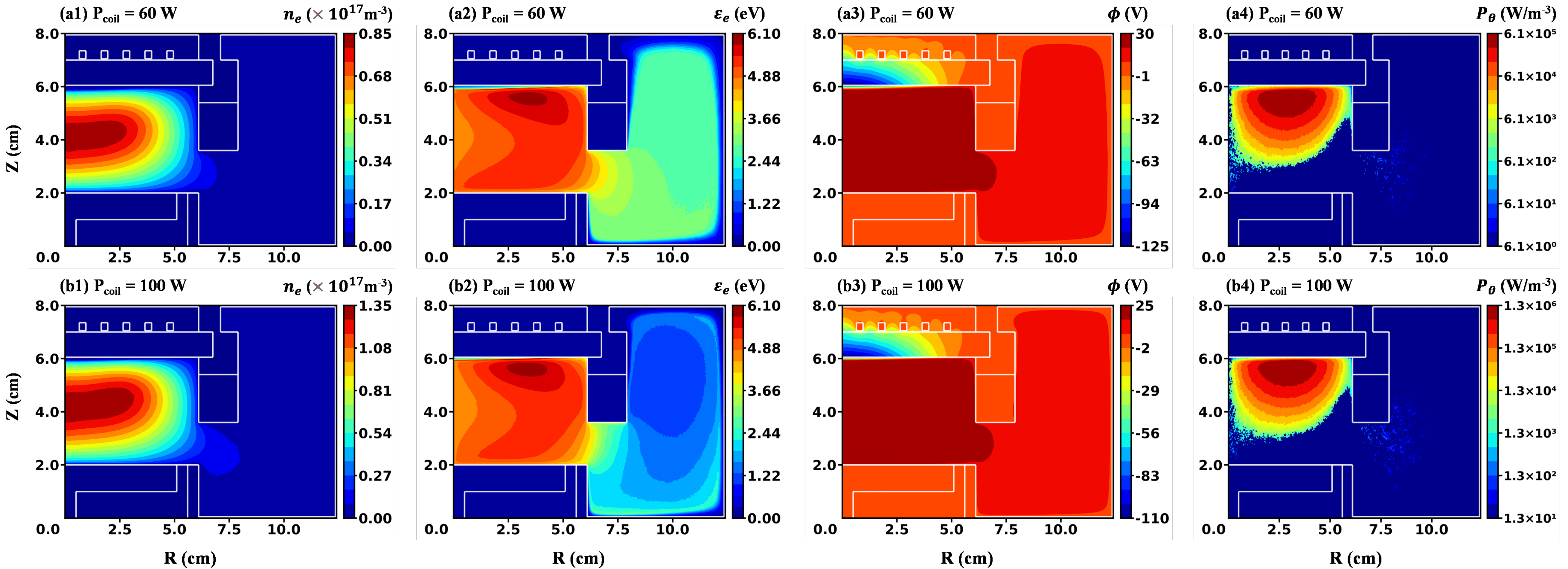}
\caption{Self-consistently calculated plasma quantities at target loaded
port powers of 60~W (top row) and 100~W (bottom row). From left to right:
electron density $n_e$, mean electron kinetic energy $\bar\varepsilon_e$,
electrostatic potential $\phi$, and RF-cycle-averaged inductive power density
$P_\theta=\langle J_\theta E_\theta\rangle$. The $n_e$,
$\bar\varepsilon_e$ and $\phi$ panels are averages over cycles 1901--2000;
the $P_\theta$ panel is the completed-cycle result at cycle 2000 and uses a
logarithmic color scale.}
\label{fig:GECplasma}
\end{figure}

Figure~\ref{fig:GECplasma} shows the spatial plasma response at the two
loaded operating points. The electron density, mean electron kinetic energy
and electrostatic potential retain similar spatial structures at the two
target powers. The displayed peak electron density rises from
$0.85\times10^{17}$ to $1.35\times10^{17}$~m$^{-3}$, while the upper plotted
value of the RF-cycle-averaged inductive power density increases from
$6.1\times10^{5}$ to $1.3\times10^{6}$~W~m$^{-3}$. The localized
$J_\theta E_\theta$ deposition beneath the coil is consistent with the
inductive heating channel, while the potential distribution also reflects
the turn-resolved capacitive coupling included in the same PIC/MCC solve.

\begin{figure*}[!htbp]
\centering
\includegraphics[width=\textwidth]{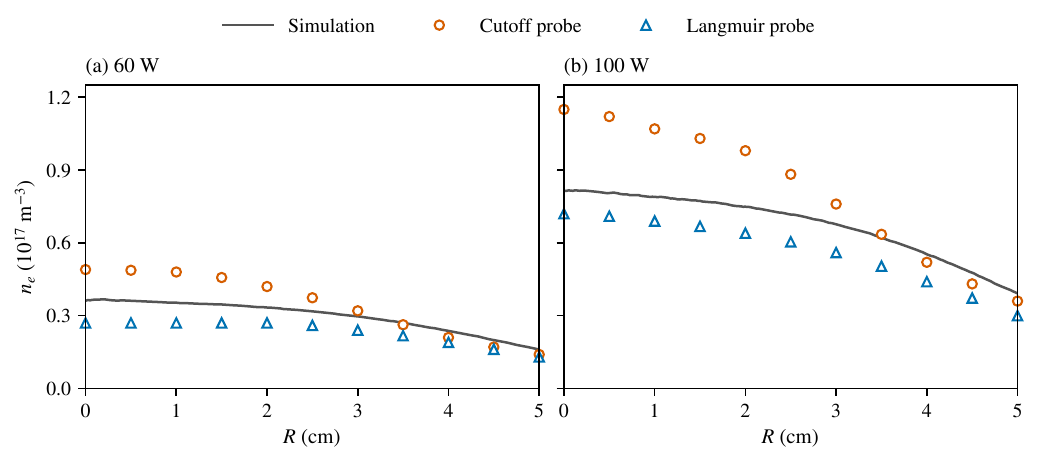}
\caption{Radial electron-density profiles from the present simulation and
the cutoff- and Langmuir-probe measurements of Sobolewski and
Kim~\cite{sobolewski2007effects}: (a)~60~W and (b)~100~W, without applied RF
bias. The panels pair simulations at controlled coil-port powers with
experiments carrying the same nominal power labels. Solid gray curves denote
the simulation; open circles and triangles
denote the cutoff- and Langmuir-probe data, respectively. Both panels use
the same density scale and display the experimental comparison interval
$0\le R\le5$~cm.}
\label{fig:GECne}
\end{figure*}

Figure~\ref{fig:GECne} compares the simulated radial electron-density
profiles with the cutoff- and Langmuir-probe measurements reported by
Sobolewski and Kim~\cite{sobolewski2007effects} under unbiased conditions at
nominal generator powers of 60 and 100~W. These experimental wattages differ
in definition from the controlled coil-terminal power $P_{\mathrm{port}}$.
At the corresponding nominal power labels, the simulation reproduces the
measured density scale and the monotonic radial decrease. The calculated
profile lies within the spread of the two probe diagnostics over most of the
displayed radius, and all three data sets show a higher density at 100~W than
at 60~W. The comparison therefore assesses the density magnitude, radial
nonuniformity, and density increase at the same nominal power labels.
It complements, rather than replaces, the analytical numerical-verification
tests in Secs.~\ref{subsec:verif_solenoid}--\ref{subsec:verif_joint_lc}.

\FloatBarrier

\begin{figure*}[!htbp]
\centering
\includegraphics[width=\textwidth]{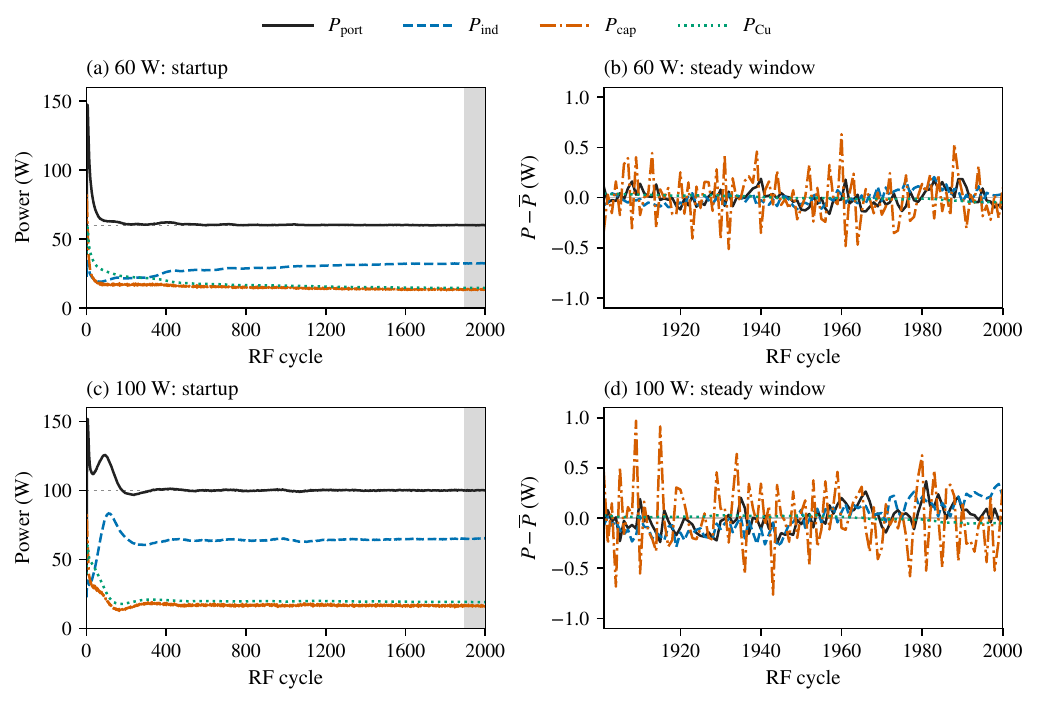}
\caption{RF-cycle evolution of the GEC coil power at target port powers of
60~W (top row) and 100~W (bottom row). Left: completed-cycle total port,
inductive, capacitive and copper-loss powers; horizontal gray dotted lines
mark the targets and shaded bands mark cycles 1901--2000. Right: deviations
of each power from its own mean over that final window, using the same line
styles. Curves show every recorded cycle without smoothing. The transient
capacitive curve represents periodically closed port work; its steady-state
heating interpretation follows Eq.~\eqref{eq:pcap_heating}.}
\label{fig:GECpowerconvergence}
\end{figure*}

Figure~\ref{fig:GECpowerconvergence} resolves the approach to the two
controlled-power operating points into total coil-port power
$P_{\mathrm{port}}$, inductive power $P_{\mathrm{ind}}$, capacitive port work
$P_{\mathrm{cap}}$, and copper resistance loss $P_{\mathrm{Cu}}$.
The initial port-power overshoot is followed by a slower redistribution
among the three power channels, particularly at 60~W. Reaching the target
port power therefore precedes the settling of the individual contributions.
All four quantities refer to the same completed RF cycle. During the
transient, $P_{\mathrm{cap}}$ is the periodically closed
$\oint V_k\,\mathrm{d}Q_k/T_{\mathrm{RF}}$ diagnostic; its identification
with net capacitive plasma heating requires the stationary conditions of
Eq.~\eqref{eq:pcap_heating}.

To quantify the power-settling time retrospectively, we identify the first
RF cycle from which
$|P_{\mathrm{port}}-P_{\mathrm{target}}|/P_{\mathrm{target}}\le1\%$
holds for all remaining recorded cycles through cycle 2000. This occurs at
cycle 737 (54.4~$\mu$s) for 60~W and cycle 1073 (79.1~$\mu$s) for 100~W.
These times characterize regulation of the total port power; the slower
redistribution among power channels remains visible in
Fig.~\ref{fig:GECpowerconvergence}.

Over cycles 1901--2000, the mean port powers are 60.07094 and 100.07067~W,
respectively, exceeding their targets by 0.118\% and 0.071\%.
The cycle-to-cycle sample standard deviations of
$(P_{\mathrm{port}},P_{\mathrm{ind}},P_{\mathrm{cap}},P_{\mathrm{Cu}})$ are
$(0.082,0.067,0.237,0.023)$~W at 60~W and
$(0.123,0.153,0.327,0.026)$~W at 100~W.
The capacitive diagnostic has the largest absolute scatter at both powers.
As a descriptive check of residual evolution, splitting this window into
cycles 1901--1950 and 1951--2000 gives changes in the four mean powers below
0.38\% of their respective full-window means. These statistics characterize
the final circuit and power window; they do not constitute an independent
test of stationarity of the particle distributions.

\begin{figure*}[!htbp]
\centering
\includegraphics[width=\textwidth]{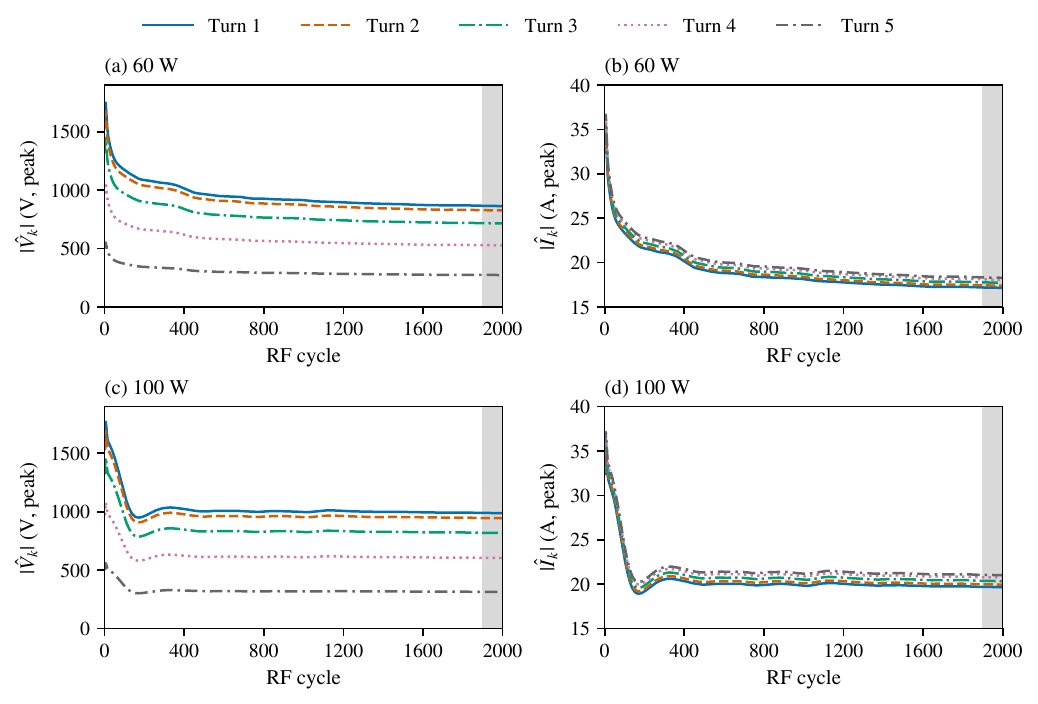}
\caption{Evolution of the five turn-potential magnitudes $|\hat V_k|$
(left) and series-current magnitudes $|\hat I_k|$ (right) at 60~W (top)
and 100~W (bottom). Turn indices follow the electrical feed-to-ground order.
All curves use peak phasors applied during cycles 2--2000; the first
available turn-resolved record is the state applied in cycle 2. Shading
marks the averaging window, cycles 1901--2000. The turn colors and line
styles are shared by all four panels; no temporal smoothing is applied.}
\label{fig:GECturnevolution}
\end{figure*}

The corresponding turn potentials and series currents evolve on the same
slow timescale, as shown in Fig.~\ref{fig:GECturnevolution}. Here
$\hat V_k$ is the potential of turn $k$ relative to ground and
$\hat I_k$ is its series-segment current. The plotted amplitudes are peak
phasor magnitudes, indexed by the RF cycle in which they were applied.
The potential decreases from the feed-side turn toward the grounded end,
whereas the series-current magnitude increases along the five turns.
The final-window changes between the two 50-cycle halves remain below
0.14\% for every turn-voltage and turn-current mean magnitude.

Together, Figs.~\ref{fig:GECpowerconvergence}
and~\ref{fig:GECturnevolution} support the feasibility of the filtered,
trend-compensated power feedback and relaxed circuit update described in
Eqs.~\eqref{eq:pfilter}--\eqref{eq:relaxation} for the two loaded GEC
cases. The strategy brings the port power into a sustained 1\% target band
on a timescale of order $10^3$ RF cycles, followed by slower relaxation of
the turn amplitudes. The final-window mean power errors below 0.12\%,
together with the small changes in every turn-amplitude mean, demonstrate
accurate power regulation and limited residual evolution of the monitored
circuit quantities over the final 100 cycles.

\begin{figure*}[!htbp]
\centering
\includegraphics[width=\textwidth]{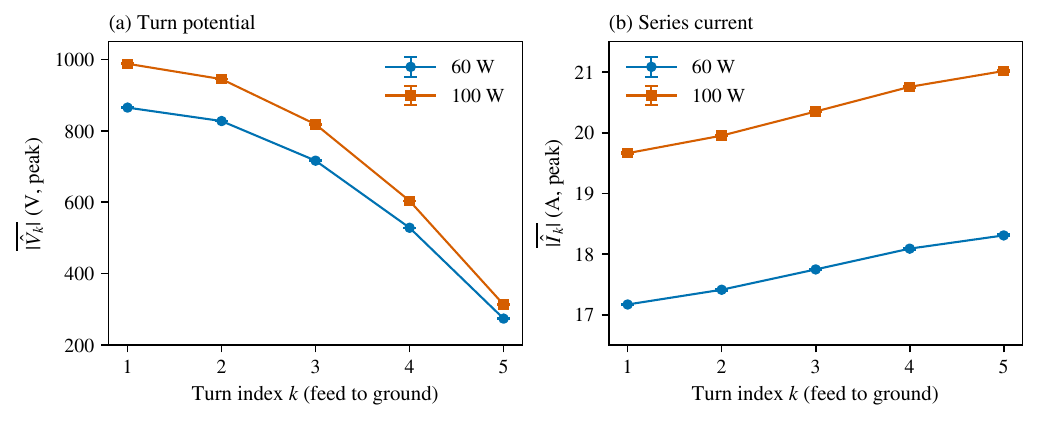}
\caption{Steady-window comparison of (a)~turn-potential and
(b)~series-current peak magnitudes at 60 and 100~W. Symbols denote means
of the cyclewise magnitudes over cycles 1901--2000; connecting lines guide
the eye. Error bars show one cycle-to-cycle sample standard deviation
(mostly smaller than the symbols), not uncertainty of the mean. The
terminal after turn 5 is grounded; its zero potential is not a sixth turn.}
\label{fig:GECturnsteady}
\end{figure*}

Figure~\ref{fig:GECturnsteady} compares the final turn distributions at the
two powers. At 60~W, the mean potential magnitude decreases from 865.37~V
on turn 1 to 273.96~V on turn 5, while the mean series-current magnitude
increases from 17.168 to 18.307~A. At 100~W, the corresponding ranges are
988.24 to 313.29~V and 19.662 to 21.017~A. Thus the turn-5 current magnitude
exceeds the turn-1 value by 6.63\% and 6.89\%, respectively. The
same-cycle phasors over cycles 1901--2000 explain this increase: at both
powers, the series current lags the local turn potential by approximately
$90^\circ$, whereas the capacitive shunt current leads it by approximately
$88^\circ$. The capacitive current is therefore nearly opposite in phase
to the incoming series current. With $C_{\mathrm{stray},k}=0$ in these
cases, Eq.~\eqref{eq:kcl} gives
$\hat I_k=\hat I_{k-1}-\hat I_{\mathrm{cap},k}$ for $k>1$;
subtracting the nearly opposing capacitive-current phasor increases the
downstream series-current magnitude. This behavior illustrates the
current redistribution produced by the coupled inductive and capacitive
responses under the present operating conditions. Conductor resistance
causes voltage drop and Joule dissipation; it does not require the current
magnitude to decrease along the winding. In the absence of all shunt
currents, KCL would instead impose equal series currents despite the
finite resistance. The nonzero potential of turn 5 is distinct from the
grounded terminal, $\hat V_6=0$;
$\hat V_k$ is not the series voltage drop $\hat V_k-\hat V_{k+1}$.

\FloatBarrier

The converged coil-port impedance was evaluated cycle by cycle as

\begin{equation}
Z_{\mathrm{port}}=\frac{\hat V_1}{\hat I_f}
\label{eq:GECzport}
\end{equation}

where $\hat V_1$ and $\hat I_f$ are the first-turn terminal voltage and feed
current of the same completed cycle. Table~\ref{tab:GECimpedance} compares
the averaged vacuum and loaded values. Plasma loading increases
$\operatorname{Re}(Z_{\mathrm{port}})$ from 0.09621085~$\Omega$ in vacuum to
0.41291192 and 0.52469544~$\Omega$ at 60 and 100~W, respectively, whereas
$\operatorname{Im}(Z_{\mathrm{port}})$ remains close to 50~$\Omega$.
The loaded impedance follows from the field--circuit solution rather than
from a fitted lumped plasma resistance. The tabulated entries are window
means only; no cycle-to-cycle standard deviations are reported with them,
and they should not be interpreted as uncertainty estimates.

\begin{table}[!htbp]
\centering
\small
\setlength{\tabcolsep}{3pt}
\caption{Coil-port impedance in vacuum and at the two loaded operating
points. Loaded values are cyclewise ratios averaged over cycles 1901--2000;
the vacuum value is averaged over completed cycles 11--15.}
\label{tab:GECimpedance}
\begin{tabular}{lccc}
\toprule
quantity & vacuum & 60~W & 100~W\\
\midrule
$\operatorname{Re}(Z_{\mathrm{port}})$ [$\Omega$]
& 0.09621085 & 0.41291192 & 0.52469544\\
$\operatorname{Im}(Z_{\mathrm{port}})$ [$\Omega$]
& 50.11822655 & 50.73066863 & 50.59666073\\
\bottomrule
\end{tabular}
\end{table}

\FloatBarrier

\begin{figure*}[!htbp]
\centering
\includegraphics[width=\textwidth]{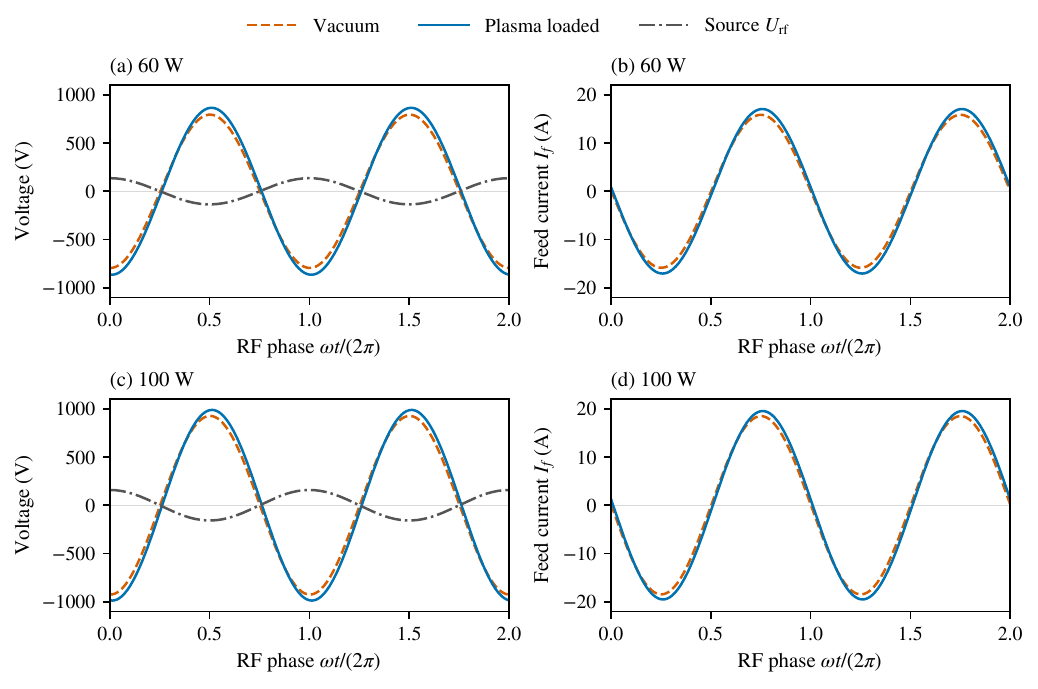}
\caption{Steady port waveforms reconstructed from peak phasors for the
60~W (top row) and 100~W (bottom row) loaded operating points. Left column:
generator voltage and the vacuum/loaded first-turn terminal voltage; right
column: the corresponding vacuum/loaded feed current. The horizontal axis
is $\omega t/(2\pi)$. Gray dash-dotted curves show the generator voltage
$U_{\mathrm{rf}}$; orange dashed and blue solid curves show the vacuum
and plasma-loaded responses, respectively. All voltages in the left column
use one common scale. The power labels identify the target loaded port
powers; the plotted terminal voltage and feed current correspond to
$\hat V_1$ and $\hat I_f$, respectively.}
\label{fig:GECui}
\end{figure*}

The corresponding steady fundamental-frequency waveforms are shown in
Fig.~\ref{fig:GECui}. At a fixed generator amplitude, plasma loading changes
both the coil-terminal voltage and feed-current phasors relative to vacuum.
The sinusoidal traces are reconstructions from the converged peak phasors.
The vacuum traces use the same generator amplitudes as their corresponding
loaded states and thereby provide same-source baselines for the loading-
induced voltage and current changes; their port powers are not assigned the
60 and 100~W loaded values.

\FloatBarrier

Finally, Table~\ref{tab:GECpower} compares the independently accumulated
inductive plasma power, capacitive port work, and conductor loss with the
circuit-side port power. In these statistically stationary, unbiased cases,
the mean rate of change of the electrostatic energy, including dielectric
and sheath storage, vanishes.
Equation~\eqref{eq:pcap_heating} therefore identifies $P_{\mathrm{cap}}$ with
net capacitive plasma heating, within numerical and sampling error. For
compactness, define

\begin{equation}
\begin{aligned}
P_{\mathrm{plasma}}&=P_{\mathrm{ind}}+P_{\mathrm{cap}},\\
P_{\mathrm{sum}}&=P_{\mathrm{plasma}}+P_{\mathrm{Cu}},\\
\Delta P&=P_{\mathrm{sum}}-P_{\mathrm{port}}
\end{aligned}
\label{eq:GECpowersum}
\end{equation}

The controlled quantity in Table~\ref{tab:GECpower} is the coil-terminal
active power, $P_{\mathrm{port}}\approx 60$ and $100~\mathrm{W}$
[Eq.~\eqref{eq:pport}]. The turn-resolved conductor loss is
$P_{\mathrm{Cu}}\approx 14.5$ and $19.1~\mathrm{W}$, while the combined
plasma heating is
$P_{\mathrm{plasma}}=P_{\mathrm{ind}}+P_{\mathrm{cap}}\approx 45.8$
and $81.4~\mathrm{W}$. These power quantities differ in definition from the
``nominal'' 60 and 100~W labels of Sobolewski and
Kim~\cite{sobolewski2007effects}, which refer to generator or matcher input
rather than $P_{\mathrm{port}}$.

The capacitive fraction $P_{\mathrm{cap}}/(P_{\mathrm{ind}}+P_{\mathrm{cap}})$
is 29.35\% at the 60~W coil-port operating point and 20.07\% at 100~W.
Its denominator is the combined plasma heating power, excluding copper
loss. This partition is specific to the present unshielded window-and-coil
geometry, rather than a universal ICP result; a Faraday shield would be
expected to reduce $P_{\mathrm{cap}}$ substantially. In the present
unshielded configuration, the capacitive contribution is not negligible at
either operating point. After
copper loss is included, the signed differences are +0.24581 and +0.50582~W,
or +0.409\% and +0.505\% of $P_{\mathrm{port}}$. The corresponding absolute
imbalances satisfy Eq.~\eqref{eq:power_imbalance}. This sub-percent closure
is quantitative internal verification that the independently implemented
field-side deposition, surface-charge work, copper loss, and circuit-port
power are mutually consistent at the converged states. It is not a separate
measurement of the particle energy lost through collisions and surfaces,
and it does not by itself equate $P_{\mathrm{port}}$ with the experimental
generator wattage.

\begin{table*}[!tbp]
\centering
\caption{Power balance for the converged loaded GEC cases. Values are means
over cycles 1901--2000. Under the steady-state, lossless-dielectric conditions
of Eq.~\eqref{eq:pcap_heating}, the capacitive port work represents net
capacitive plasma heating. The displayed sums and differences are computed
from the unrounded source values.}
\label{tab:GECpower}
\small
\begin{tabular}{@{}p{3.4cm}p{5.5cm}rr@{}}
\toprule
quantity & definition or comparison basis & 60~W & 100~W\\
& & [W] & [W]\\
\midrule
$P_{\mathrm{ind}}$ & volume integral of $J_\theta E_\theta$
& 32.33719 & 65.08536\\
$P_{\mathrm{cap}}$ & sum of $T_{\mathrm{RF}}^{-1}\oint V_k\,\mathrm{d}Q_k$
over the turns & 13.43354 & 16.34652\\
$P_{\mathrm{plasma}}$ & $P_{\mathrm{ind}}+P_{\mathrm{cap}}$
& 45.77073 & 81.43188\\
$P_{\mathrm{Cu}}$ & turn-resolved RF conductor loss
& 14.54603 & 19.14460\\
$P_{\mathrm{sum}}$ & $P_{\mathrm{plasma}}+P_{\mathrm{Cu}}$
& 60.31676 & 100.57649\\
$P_{\mathrm{port}}$ & $\tfrac12\operatorname{Re}(\hat V_1\hat I_f^*)$
& 60.07094 & 100.07067\\
\midrule
$\Delta P$ & $P_{\mathrm{sum}}-P_{\mathrm{port}}$
& $+0.24581$ & $+0.50582$\\
$\Delta P/P_{\mathrm{port}}$ & signed relative difference
& $+0.409\%$ & $+0.505\%$\\
\bottomrule
\end{tabular}
\end{table*}

The numerical assessment focuses on the added field--circuit coupling:
Tests~1--3 examine the field-derived port responses and network closure,
whereas the GEC case exercises the algorithm with a self-consistent kinetic
plasma. A systematic assessment of the underlying PIC discretization and
particle-sampling uncertainty is outside the scope of this work.

Taken together, the GEC results show that the bidirectionally coupled model
reaches both controlled coil-port operating points while resolving
substantial inductive and capacitive plasma heating in an unshielded
geometry. The density profiles provide an experimental, profile-level
comparison at the same nominal power labels, whereas the loading-induced
impedance shift and sub-percent power closure provide complementary evidence
for the coupled circuit response and internal energy consistency.

\FloatBarrier

\section{Conclusions}
\label{sec:conclusions}

This work developed a self-consistent phasor--PIC algorithm that couples a
distributed RF coil circuit to an axisymmetric ICP through both inductive and
capacitive feedback. The field-derived unit responses supply the vacuum
self- and mutual-inductive impedance matrix, the plasma-only azimuthal field
supplies the per-turn series back-emfs, and the Poisson-consistent conductor
charges supply the per-turn shunt displacement currents. A distributed
$(2N+1)$ complex network then updates the coil currents and turn potentials
once per RF period, thereby allowing the plasma load to modify the circuit
state used by the subsequent PIC/MCC cycle.

Verification covered the component inductive responses and their joint
closure with capacitive feedback. For the single-solenoid test, the base
field-derived inductance differed from the infinite-solenoid value by
$-1.58$\%, and all seven boundary, mesh, and length cases remained within
$3$\% in magnitude. In the two-solenoid test, the mutual inductance differed
from the analytical reference by $-4.38$\%, while the passive-secondary
voltage and current differed by $4.730$\%. The five-node manufactured test
then exercised the field-derived inductive matrix, numerical Laplace solve,
surface-charge reduction, and eleven-unknown network with both channels
active. At a 1-mm spacing, the scaled complex-state errors were
$1.90\times10^{-5}$ against the continuous reference and
$6.26\times10^{-12}$ against the independently assembled semidiscrete
network. Four-grid refinement gave approximately second-order convergence,
and the computed period-delay history agreed with the independent recurrence
to $7.40\times10^{-12}$. These results verify the simultaneous linear
field--circuit coupling in the particle-free benchmark. On the principal grid, the reciprocity,
capacitive-power, and coil-loss errors are $6.27\times10^{-6}$,
$1.004\times10^{-6}$, and $4.51\times10^{-7}$, respectively.
The test establishes numerical accuracy within these reported errors,
without implying exact finite-grid losslessness or verification
of nonlinear plasma feedback and particle heating.

Application to the argon GEC reference cell combined both feedback channels
with the kinetic plasma calculation. The simulated density profiles were
consistent with the measured density scale, the monotonic radial decrease,
and the density increase from 60 to 100~W. The experimental wattages are
nominal generator values and are not identical to the controlled
coil-terminal power $P_{\mathrm{port}}$. Plasma loading increased the real
part of the coil-port impedance from
0.0962~$\Omega$ in vacuum to 0.413 and 0.525~$\Omega$ at the two
loaded operating points. At these statistically stationary states, the
capacitive port work represents net capacitive plasma heating because the
field energy has no mean increment and the dielectric has no dissipative
loss. At coil-terminal powers of 60 and 100~W, conductor loss accounted for
about 14.5 and 19.1~W, respectively, with combined plasma heating of
$P_{\mathrm{plasma}}\approx 45.8$ and $81.4~\mathrm{W}$. Capacitive heating
contributed 29.35\% and 20.07\% of $P_{\mathrm{ind}}+P_{\mathrm{cap}}$
in this unshielded window-and-coil geometry, with copper loss excluded from
the denominator. These fractions are not claimed for Faraday-shielded
antennas. The independently accumulated
field-side deposition, conductor loss, and circuit-port power closed with
relative imbalances of 0.409\% and 0.505\%, providing an internal
energy-consistency check of the converged coupled states.

The calculations reported here use an axisymmetric, fundamental-frequency
representation, a local skin-effect resistance model, and a fixed
single-ended coil topology. The partitioned
field--circuit update also contains an explicit one-RF-period delay, and
capacitances outside the resolved Poisson domain must be supplied as external
parameters. On the verification side, the analytical references cover the
particle-free magnetoquasistatic limit; an independent reference case with a
prescribed nonzero plasma back-emf would further exercise the inductive
feedback path and remains future work.

The frequency-domain circuit equations provide a direct route to a
multiharmonic extension: the same network structure can be solved at each
retained frequency $m\omega$, where $m$ is the harmonic order, using the
corresponding field responses and frequency-dependent conductor and
matching-network impedances. Plasma-current and conductor-charge harmonics
would supply the feedback at each frequency, with
$\hat I_{\mathrm{cap},k}^{(m)}=\mathrm{i}m\omega\hat Q_k^{(m)}$.
Reconstructing the combined waveforms for the time-resolved PIC advance
would retain the nonlinear coupling between harmonics through the plasma
response. This extension would support self-consistent circuit studies of
CWT, building on our earlier work with prescribed multiharmonic coil
currents~\cite{chen2026cwt}. For larger planar coils, a refined conductor-loss
model could account for winding-dependent skin and proximity effects beyond
the present local estimate. The self-consistent loaded impedance could also
guide optimal impedance-matching design by varying matching-network
capacitances and inductances to reduce reflected power, subject to limits on
turn voltages and conductor losses. A three-dimensional extension of the
field solves and conductor-charge extraction would enable studies of
non-axisymmetric plasma structure and realistic helical-coil and feed
geometries. Subcycle or fully implicit field--circuit coupling is a further
direction for improving the treatment of the delayed plasma response.


\section*{Declaration of competing interest}
The authors declare that they have no known competing financial interests or
personal relationships that could have appeared to influence the work
reported in this paper.

\section*{Code availability}
The PIC/MCC source code used in this study is closed-source and is not
distributed. The mathematical formulation and numerical coupling algorithm
are described in this article.

\section*{Data availability}
The numerical data supporting the findings of this study are available
from the corresponding authors upon reasonable request.

\section*{Declaration of generative AI and AI-assisted technologies in the manuscript preparation process}
During the preparation of this work, the authors used OpenAI Codex to
improve the language expression, readability, and LaTeX formatting of the
manuscript. The authors reviewed and verified all AI-assisted content and
take full responsibility for the final manuscript.

\section*{Acknowledgements}
This work was supported by the National Magnetic Confinement Fusion Energy
Research Project (2022YFE03190300) and the National Natural Science Foundation
of China (12675259, 12675325, and W2621001).

\appendix
\setcounter{figure}{0}
\setcounter{table}{0}
\renewcommand{\theHfigure}{appendix.\arabic{figure}}
\renewcommand{\theHtable}{appendix.\arabic{table}}
\section{Additional checks for the joint benchmark}
\label{app:joint_lc_details}

\begin{figure*}[!htbp]
\centering
\includegraphics[width=\textwidth]{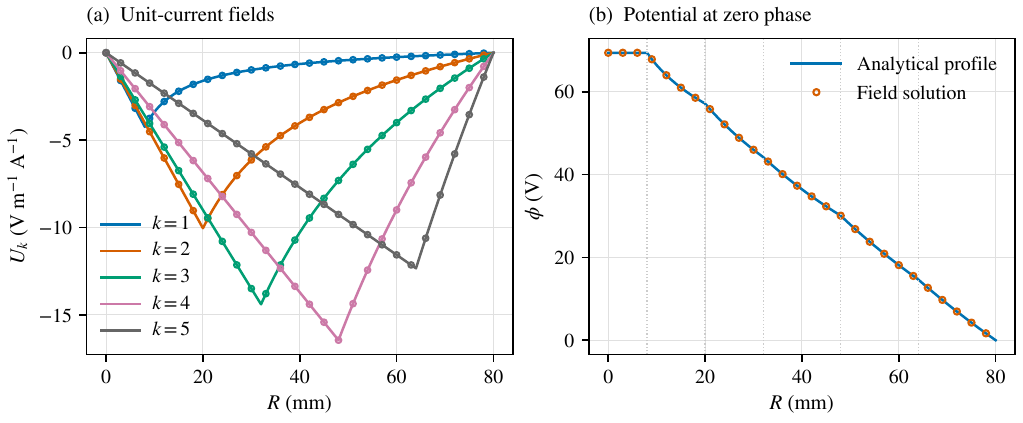}
\caption{Midplane field profiles for the principal grid, $h=1$~mm.
(a)~Five unit-current inductive coefficients $U_k(R)$; each color identifies
one source sheet. (b)~Electrostatic potential at zero RF phase at the end
of cycle~30. The analytical potential uses the voltages actually applied
in that completed cycle. In both panels, lines are analytical solutions
and open symbols sample the numerical field every third radial point for
legibility. Dotted vertical lines in (b) mark the sheet radii.}
\label{fig:t7fields}
\end{figure*}

The radial solutions specify the manufactured axial boundary conditions
and provide local field checks in addition to the integrated network
comparisons. The real unit-current inductive coefficient is
\begin{equation}
U_k(R)=-\frac{\omega\mu_0}{2\ell}
\begin{cases}
(1-a_k^2/b^2)R, & 0\le R\le a_k,\\
a_k^2(1/R-R/b^2), & a_k<R\le b
\end{cases}
\label{eq:t7unitfield}
\end{equation}
It has units of V\,m$^{-1}$\,A$^{-1}$ and gives
$\hat E_\theta=\mathrm{i}\sum_k\hat I_kU_k$ under the stated peak-phasor
convention. Its current-sheet projection gives the finite-$b$ matrix in
Eq.~\eqref{eq:t7M}. The electrostatic solution is $\hat\phi=\hat V_1$
inside the innermost sheet and, for $a_k<R<a_{k+1}$,
\begin{equation}
\hat\phi(R)=
\frac{\hat V_k\ln(a_{k+1}/R)+\hat V_{k+1}\ln(R/a_k)}
     {\ln(a_{k+1}/a_k)},\qquad
\hat E_R=\frac{\hat V_k-\hat V_{k+1}}
{R\ln(a_{k+1}/a_k)}
\label{eq:t7potential}
\end{equation}
with $a_6=b$ and $\hat V_6=0$. The charge convention is
$\hat Q_k=\int_{\partial\Omega_k}\varepsilon
\hat{\boldsymbol E}\cdot\boldsymbol n_k\,\mathrm dS$, where
$\boldsymbol n_k$ points from the conductor into the surrounding domain.
It yields $\hat{\mathbf Q}=\mathbf C\hat{\mathbf V}$.
Figure~\ref{fig:t7fields} compares the numerical and analytical radial
profiles, and Table~\ref{tab:t7phasors} lists the final network phasors.

The field errors use the cylindrical weighted norm
\begin{equation}
\epsilon_{2,R}(u)=
\left[\frac{\sum_{j\in\mathcal J}R_j|u_{h,j}-u_j|^2}
{\sum_{j\in\mathcal J}R_j|u_j|^2}\right]^{1/2}
\label{eq:t7fieldnorm}
\end{equation}
For $U_k$, the comparison uses the central axial plane and radial points
with nonzero reference value. The largest error among the five fields is
$1.02\times10^{-3}$. The potential error is $8.32\times10^{-6}$, using
points where the reference exceeds $10^{-6}$ of its maximum magnitude.
For the radial electric field, the numerical profile uses centered
potential differences in the interior and a one-sided difference at the
outer boundary; the axis and points within $2h$ of a sheet
are excluded, together with reference values below the same relative
cutoff. Its error is $1.07\times10^{-3}$ and does not describe accuracy
at the derivative jumps. These are deterministic comparisons without
statistical error bars.

Although the continuous manufactured solution is independent of $z$, the
discrete field has small axial boundary layers. At $h=1$~mm the maximum
departure from the central profile near either end, normalized by the
maximum analytical unit field, is $1.95\times10^{-3}$. It falls to
$1.23\times10^{-4}$ at $h=0.25$~mm. The maximum departure within the
central axial window (approximately 10\% of the interior rows) is
$6.61\times10^{-7}$ at $h=1$~mm. Consequently, axial uniformity is a
continuum property of the manufactured reference, not an exact property
of the finite-grid solution.

\begin{table*}[!tbp]
\centering
\caption{Final complex peak phasors after 30 RF periods at $h=1$~mm,
compared with the continuous solution of Eq.~\eqref{eq:t7network}.
Values are rounded to eight decimal places to resolve the numerical
differences. Relative complex-phasor errors against the continuous and
semidiscrete references are shown in Fig.~\ref{fig:t7accuracy}(d).}
\label{tab:t7phasors}
\small
\begin{tabular}{@{}llrr@{}}
\toprule
Unknown & Unit & Field--circuit calculation & Continuous reference \\
\midrule
$\hat I_f$ & A & $0.61201598-0.02550914\mathrm{i}$ & $0.61201046-0.02552834\mathrm{i}$ \\
$\hat I_1$ & A & $0.60664272-0.08266655\mathrm{i}$ & $0.60664332-0.08265052\mathrm{i}$ \\
$\hat I_2$ & A & $0.60823421-0.14284113\mathrm{i}$ & $0.60823228-0.14284154\mathrm{i}$ \\
$\hat I_3$ & A & $0.62552190-0.17418067\mathrm{i}$ & $0.62551742-0.17418805\mathrm{i}$ \\
$\hat I_4$ & A & $0.63827910-0.25383117\mathrm{i}$ & $0.63827332-0.25384067\mathrm{i}$ \\
$\hat I_5$ & A & $0.60801444-0.30612588\mathrm{i}$ & $0.60800732-0.30613602\mathrm{i}$ \\
$\hat V_1$ & V & $69.39920116+1.27545705\mathrm{i}$ & $69.39947722+1.27641707\mathrm{i}$ \\
$\hat V_2$ & V & $57.12975940+2.42888340\mathrm{i}$ & $57.12997928+2.42924426\mathrm{i}$ \\
$\hat V_3$ & V & $44.20463556+2.84545576\mathrm{i}$ & $44.20479252+2.84551168\mathrm{i}$ \\
$\hat V_4$ & V & $30.07491397+1.56178999\mathrm{i}$ & $30.07500916+1.56170406\mathrm{i}$ \\
$\hat V_5$ & V & $14.67817659-0.20929854\mathrm{i}$ & $14.67819416-0.20939838\mathrm{i}$ \\
\bottomrule
\end{tabular}
\end{table*}

The auxiliary perturbations retain the same geometry and source frequency.
A $37^\circ$ rotation of the source rotates the solution by the same
phasor factor, with a normalized discrepancy of
$7.08\times10^{-15}$. Setting all conductor potentials to zero gives zero
conductor charge and displacement current in the capacitive calculation.

For the MPI, phase-sampling, and solver-tolerance comparisons below, the
state metric is $\max_j|x_j^{(a)}-x_j^{(b)}|/\max_j|x_j^{(b)}|$, evaluated
using numerical current and voltage entries in amperes and volts. This
metric depends on the chosen units and differs from the dimensionless
scaled norm in Eq.~\eqref{eq:t7error}. Relaxing the GMRES relative tolerance
from $10^{-12}$ to $10^{-5}$ changes the final state by
$3.67\times10^{-5}$ under this metric. The semidiscrete-reference accuracy
reported in Sec.~\ref{subsec:verif_joint_lc} therefore applies to the
$10^{-12}$ tolerance.

Computations with 1, 2, 4, and 8 MPI ranks, including separate two-rank
axial and radial decompositions, give a maximum state difference
of $1.32\times10^{-11}$. Increasing the phase samples from 400 to 800
gives a difference of $1.44\times10^{-15}$. A 15-period run followed by
a 15-period restart gives identical diagnostics for cycles 16--30 to
those of an uninterrupted 30-period calculation. These comparisons assess
numerical consistency across parallel decompositions, phase resolutions,
and restart conditions. The calculations used Intel Fortran 2021.9.0 and
Intel MPI 2021.9.0.

The tested electrical ordering is the identity order $1,\ldots,5$;
arbitrary network permutations are outside the present evidence.

After examining the finite-grid errors, the tolerance for $\eta_M$ was revised
from $10^{-6}$ to $10^{-4}$, that for $\eta_{\mathrm{cap}}$ from
$10^{-6}$ to $10^{-5}$, and that for $\eta_{\mathrm{coil}}$ from
$10^{-8}$ to $10^{-5}$. The same revised bounds apply to all four grids.
The computed errors in Table~\ref{tab:t7defects} are unaffected by this
change in tolerances.
\FloatBarrier

\bibliographystyle{elsarticle-num}
\bibliography{ref}

@article{sobolewski2007effects,
	title = {The effects of radio-frequency bias on electron density in an inductively coupled plasma reactor},
	author = {Sobolewski, Mark A and Kim, Jung-Hyung},
	journal = {Journal of Applied Physics},
	volume = {102},
	number = {11},
	pages = {113302},
	doi = {10.1063/1.2815674},
	year = {2007},
	publisher = {AIP Publishing},
}

@article{xu2015equivalent,
	author = {Xu, Hui-Jing and Zhao, Shu-Xia and Zhang, Yu-Ru and Gao, Fei and Li, Xue-Chun and Wang, You-Nian},
	title = {Equivalent circuit effects on mode transitions in H2 inductively coupled plasmas},
	journal = {Physics of Plasmas},
	volume = {22},
	pages = {043508},
	doi = {10.1063/1.4917335},
	year = {2015},
	publisher = {AIP Publishing},
}

@article{sydorenko2025darwin,
	author = {Sydorenko, Dmytro and Kaganovich, Igor D. and Khrabrov, Alexander V. and Ethier, Stephane A. and Chen, Jin and Janhunen, Salomon},
	title = {Simulation of an inductively coupled plasma with a two-dimensional Darwin particle-in-cell code},
	journal = {Physics of Plasmas},
	volume = {32},
	pages = {043904},
	doi = {10.1063/5.0241152},
	year = {2025},
	publisher = {AIP Publishing},
}

@article{singh2008electrical,
	author = {Singh, S. V. and Pargmann, C.},
	title = {Electrical characterization of an inductively coupled gaseous electronics conference reference cell},
	journal = {Journal of Applied Physics},
	volume = {104},
	pages = {083303},
	doi = {10.1063/1.3000667},
	year = {2008},
	publisher = {AIP Publishing},
}

@article{godyak2017power,
	author = {Godyak, V. A. and Alexandrovich, B. M.},
	title = {Power measurements and coupler optimization in inductive discharges},
	journal = {Review of Scientific Instruments},
	volume = {88},
	pages = {083512},
	doi = {10.1063/1.4995810},
	year = {2017},
	publisher = {AIP Publishing},
}

@article{vahedi1995mcc,
	author = {Vahedi, V. and Surendra, M.},
	title = {A Monte Carlo collision model for the particle-in-cell method: applications to argon and oxygen discharges},
	journal = {Computer Physics Communications},
	volume = {87},
	number = {1--2},
	pages = {179--198},
	doi = {10.1016/0010-4655(94)00171-W},
	year = {1995},
	publisher = {Elsevier},
}

@article{mattei2017fully,
	author = {Mattei, S. and Nishida, K. and Onai, M. and Lettry, J. and Tran, M. Q. and Hatayama, A.},
	title = {A fully-implicit Particle-In-Cell Monte Carlo Collision code for the simulation of inductively coupled plasmas},
	journal = {Journal of Computational Physics},
	volume = {350},
	pages = {891--906},
	doi = {10.1016/j.jcp.2017.09.015},
	year = {2017},
	publisher = {Elsevier},
}

@article{qu2020power,
	author = {Qu, Chenhui and Lanham, Steven J. and Shannon, Steven C. and Nam, Sang Ki and Kushner, Mark J.},
	title = {Power matching to pulsed inductively coupled plasmas},
	journal = {Journal of Applied Physics},
	volume = {127},
	pages = {133302},
	doi = {10.1063/5.0002522},
	year = {2020},
	publisher = {AIP Publishing},
}

@article{hargis1994gaseous,
	author = {Hargis, P. J. and Greenberg, K. E. and Miller, P. A. and Gerardo, J. B. and Torczynski, J. R. and Riley, M. E. and Hebner, G. A. and Roberts, J. R. and Olthoff, J. K. and Whetstone, J. R. and Van Brunt, R. J. and Sobolewski, M. A. and Anderson, H. M. and Bishop, M. P. and Kindel, E. E. and Bose, B. and Sandel, J. C. and Han, W. Y. and Nutter, M. T. and Patel, S. and Paul, T. M. and Shim, S. K. and \'O, J. K. and Ashford, D. C.},
	title = {The Gaseous Electronics Conference radio-frequency reference cell: A defined parallel-plate radio-frequency system for experimental and theoretical studies of plasma-processing discharges},
	journal = {Review of Scientific Instruments},
	volume = {65},
	pages = {140--154},
	doi = {10.1063/1.1144770},
	year = {1994},
	publisher = {AIP Publishing},
}

@article{watanabe1999radio,
	author = {Watanabe, M. and Shaw, D. M. and Collins, G. J. and Sugai, H.},
	title = {Radio-frequency plasma potential variations originating from capacitive coupling from the coil antenna in inductively coupled plasmas},
	journal = {Journal of Applied Physics},
	volume = {85},
	number = {7},
	pages = {3428--3434},
	doi = {10.1063/1.369700},
	year = {1999},
	publisher = {AIP Publishing},
}

@article{chen2024gec,
	author = {Chen, Zili and Wang, Hongyu and Yu, Shimin and Wang, Yu and Chen, Zhipeng and Jiang, Wei and Schulze, Julian and Zhang, Ya},
	title = {Electrical characteristics of the GEC reference cell at low pressure: a two-dimensional PIC/MCC modeling study},
	journal = {Plasma Sources Science and Technology},
	volume = {33},
	number = {4},
	pages = {045003},
	doi = {10.1088/1361-6595/ad3849},
	year = {2024},
	publisher = {IOP Publishing},
}

@article{chen2025eh,
	author = {Chen, Zhaoyu and Chen, Zili and Wang, Yu and Jiang, Wei and Ding, Yilin and Xia, Dong and Zhang, Ya},
	title = {Simulations of E-H mode transition in inductively coupled plasmas via 2D particle-in-cell/Monte Carlo collision method},
	journal = {Plasma Sources Science and Technology},
	volume = {34},
	pages = {095009},
	doi = {10.1088/1361-6595/ae05c6},
	year = {2025},
	publisher = {IOP Publishing},
}

@article{chen2026cwt,
	author = {Chen, Zhaoyu and Chen, Zili and Wang, Yu and Giesekus, Jonas and Jiang, Wei and Ding, Yonghua and Xia, Donghui and Zhang, Ya and Schulze, Julian},
	title = {Control of electron energy probability function by current waveform tailoring in inductively coupled radio frequency plasmas},
	journal = {Physics of Plasmas},
	volume = {33},
	pages = {093506},
	doi = {10.1063/5.0343175},
	year = {2026},
	publisher = {AIP Publishing},
}

@article{chen2022winding,
	author = {Chen, Qingbin and Zhang, Xu and Chen, Wei and Wang, Cong},
	title = {Winding loss analysis of planar spiral coil and its structure optimization technique in wireless power transfer system},
	journal = {Scientific Reports},
	volume = {12},
	pages = {19418},
	doi = {10.1038/s41598-022-24006-x},
	year = {2022},
}

@article{chen2025swpc,
	author = {Chen, Zili and Chen, Zhaoyu and Wang, Yu and Xu, Jingwen and Chen, Zhipeng and Jiang, Wei and Wang, Hongyu and Zhang, Ya},
	title = {Stochastic weighted particle control for electrostatic particle-in-cell Monte Carlo collision simulations in an axisymmetric coordinate system},
	journal = {Computer Physics Communications},
	volume = {306},
	pages = {109390},
	doi = {10.1016/j.cpc.2024.109390},
	year = {2025},
	publisher = {Elsevier},
}

@article{chen2026rf,
	author = {Chen, Zhaoyu and Chen, Zili and Wang, Yu and Wang, Hongyu and Jiang, Wei and Ding, Yilin and Xia, Dong and Schulze, Julian and Chen, Xingyu and Zhang, Ya},
	title = {Effects of RF bias phase on plasma characteristics in an inductively coupled plasma},
	journal = {Plasma Sources Science and Technology},
	volume = {35},
	pages = {025030},
	doi = {10.1088/1361-6595/ae4537},
	year = {2026},
	publisher = {IOP Publishing},
}

@article{fu2024kinetic,
	author = {Fu, Chencong and Dong, Yicheng and Li, Yifei and Wang, Weizong and Wang, Zihan and Liu, Wei},
	title = {Kinetic simulations of low-pressure inductively coupled plasma: an implicit electromagnetic PIC/MCC model with the ADI-FDTD method},
	journal = {Journal of Physics D: Applied Physics},
	volume = {57},
	pages = {135201},
	doi = {10.1088/1361-6463/ad1729},
	year = {2024},
	publisher = {IOP Publishing},
}

@article{guittienne2017electromagnetic,
	author = {Guittienne, Ph and Jacquier, R and Howling, A A and Furno, I},
	title = {Electromagnetic, complex image model of a large area RF resonant antenna as inductive plasma source},
	journal = {Plasma Sources Science and Technology},
	volume = {26},
	pages = {035010},
	doi = {10.1088/1361-6595/aa59d6},
	year = {2017},
	publisher = {IOP Publishing},
}

@article{elfayoumi1998electromagnetic,
	author = {El-Fayoumi, I M and Jones, I R},
	title = {The electromagnetic basis of the transformer model for an inductively coupled RF plasma source},
	journal = {Plasma Sources Science and Technology},
	volume = {7},
	pages = {179--185},
	doi = {10.1088/0963-0252/7/2/012},
	year = {1998},
	publisher = {IOP Publishing},
}

@article{lymberopoulos1995two,
	author = {Lymberopoulos, D P and Economou, D J},
	title = {Two-Dimensional Self-Consistent Radio Frequency Plasma Simulations Relevant to the Gaseous Electronics Conference RF Reference Cell},
	journal = {Journal of Research of the National Institute of Standards and Technology},
	volume = {100},
	number = {4},
	pages = {473--494},
	doi = {10.6028/jres.100.036},
	year = {1995},
	publisher = {NIST},
}

@article{miller1995inductively,
	author = {Miller, P A and Hebner, G A and Greenberg, K E and Pochan, P D and Aragon, B P},
	title = {An Inductively Coupled Plasma Source for the Gaseous Electronics Conference RF Reference Cell},
	journal = {Journal of Research of the National Institute of Standards and Technology},
	volume = {100},
	number = {4},
	pages = {427--439},
	doi = {10.6028/jres.100.032},
	year = {1995},
	publisher = {NIST},
}

@article{schmidt2018consistent,
	author = {Schmidt, Frederik and Mussenbrock, Thomas and Trieschmann, Jan},
	title = {Consistent simulation of capacitive radio-frequency discharges and external matching networks},
	journal = {Plasma Sources Science and Technology},
	volume = {27},
	pages = {105017},
	doi = {10.1088/1361-6595/aae429},
	year = {2018},
	publisher = {IOP Publishing},
}

@article{verboncoeur1993simultaneous,
	author = {Verboncoeur, J. P. and Alves, M. V. and Vahedi, V. and Birdsall, C. K.},
	title = {Simultaneous Potential and Circuit Solution for 1D Bounded Plasma Particle Simulation Codes},
	journal = {Journal of Computational Physics},
	volume = {104},
	pages = {321--328},
	doi = {10.1006/jcph.1993.1034},
	year = {1993},
	publisher = {Elsevier},
}

@article{ruehli1974equivalent,
	author = {Ruehli, Albert E.},
	title = {Equivalent {Circuit} {Models} for {Three}-{Dimensional} {Multiconductor} {Systems}},
	journal = {IEEE Transactions on Microwave Theory and Techniques},
	volume = {22},
	number = {3},
	pages = {216--221},
	doi = {10.1109/TMTT.1974.1128204},
	year = {1974},
	publisher = {IEEE},
}

@article{ho1975modified,
	author = {Ho, Chung-Wen and Ruehli, Albert E. and Brennan, Pierce A.},
	title = {The {Modified} {Nodal} {Approach} to {Network} {Analysis}},
	journal = {IEEE Transactions on Circuits and Systems},
	volume = {22},
	number = {6},
	pages = {504--509},
	doi = {10.1109/TCS.1975.1084079},
	year = {1975},
	publisher = {IEEE},
}

@article{brackbill1982implicit,
	author = {Brackbill, J. U. and Forslund, D. W.},
	title = {An {Implicit} {Method} for {Electromagnetic} {Plasma} {Simulation} in {Two} {Dimensions}},
	journal = {Journal of Computational Physics},
	volume = {46},
	number = {2},
	pages = {271--308},
	doi = {10.1016/0021-9991(82)90016-x},
	year = {1982},
	publisher = {Elsevier},
}

@article{hewett1994low,
	author = {Hewett, D. W.},
	title = {Low-frequency electromagnetic ({Darwin}) applications in plasma simulation},
	journal = {Computer Physics Communications},
	volume = {84},
	number = {1-3},
	pages = {243--277},
	doi = {10.1016/0010-4655(94)90214-3},
	year = {1994},
	publisher = {Elsevier},
}

@article{langdon1983direct,
	author = {Langdon, A. B. and Cohen, B. I. and Friedman, A.},
	title = {Direct {Implicit} {Large} {Time}-{Step} {Particle} {Simulation} of {Plasmas}},
	journal = {Journal of Computational Physics},
	volume = {51},
	number = {1},
	pages = {107--138},
	doi = {10.1016/0021-9991(83)90083-9},
	year = {1983},
	publisher = {Elsevier},
}

@article{angus2024axisymmetric,
	author = {Angus, Justin Ray and Farmer, William and Friedman, Alex and Geyko, Vasily and Ghosh, Debojyoti and Grote, Dave and Larson, David and Link, Anthony},
	title = {An implicit particle code with exact energy and charge conservation for studies of dense plasmas in axisymmetric geometries},
	journal = {Journal of Computational Physics},
	volume = {519},
	pages = {113427},
	doi = {10.1016/j.jcp.2024.113427},
	year = {2024},
	publisher = {Elsevier},
}

@article{na2024borpic,
	author = {Na, Dong-Yeop and Teixeira, Fernando L. and Omelchenko, Yuri A.},
	title = {An unstructured body-of-revolution electromagnetic particle-in-cell algorithm with radial perfectly matched layers and dual polarizations},
	journal = {Computer Physics Communications},
	volume = {302},
	pages = {109247},
	doi = {10.1016/j.cpc.2024.109247},
	year = {2024},
	publisher = {Elsevier},
}

@article{mao2025poisson,
	author = {Mao, Renfan and Ren, Junxue and Tang, Haibin and Li, Zhihui},
	title = {A domain decomposition parallelization scheme for the {Poisson} equation in particle-in-cell simulation},
	journal = {Computer Physics Communications},
	volume = {310},
	pages = {109527},
	doi = {10.1016/j.cpc.2025.109527},
	year = {2025},
	publisher = {Elsevier},
}

@article{eremin2022bounded,
	author = {Eremin, D.},
	title = {An energy- and charge-conserving electrostatic implicit particle-in-cell algorithm for simulations of collisional bounded plasmas},
	journal = {Journal of Computational Physics},
	volume = {452},
	pages = {110934},
	doi = {10.1016/j.jcp.2021.110934},
	year = {2022},
	publisher = {Elsevier},
}

@article{eckert2025benchmark,
	author = {Eckert, Zakari and Boerner, Jeremiah J. and Hall, Taylor H. and Hooper, Russell and Grillet, Anne M. and Pacheco, Jose L.},
	title = {Benchmark verification of {PIC-DSMC} programs},
	journal = {Journal of Computational Physics},
	volume = {521},
	pages = {113533},
	doi = {10.1016/j.jcp.2024.113533},
	year = {2025},
	publisher = {Elsevier},
}

\end{document}